\documentclass[aps,10pt,prd,twocolumn,superscriptaddress,preprintnumbers,%
showkeys,nofootinbib,showpacs]{revtex4-2}
\usepackage{slashed}
\usepackage{eucal} 
\usepackage{bm}
\usepackage{graphicx}
\usepackage[hypertexnames=false]{hyperref}
\usepackage{amsmath,amssymb,amsfonts,latexsym}
\begin{document}
\author{June-Young Kim}
\email[ E-mail: ]{jykim@jlab.org}
\affiliation{Theory Center, Jefferson Lab, Newport News, VA 23606, USA}
\author{Ho-Yeon Won}
\email[E-mail: ]{hoyeon.won@polytechnique.edu}
\affiliation{CPHT, CNRS, \'Ecole polytechnique, Institut Polytechnique de Paris,
91120 Palaiseau, France}
\author{Hyun-Chul Kim}
\email[ E-mail: ]{hchkim@inha.ac.kr}
\affiliation{Department of Physics, Inha University, Incheon 402-751, Republic of Korea} 
\affiliation{School of Physics, Korea Institute for Advanced Study (KIAS), Seoul 02455,
Republic of Korea}
\author{Christian Weiss}
\email[ E-mail: ]{weiss@jlab.org}
\affiliation{Theory Center, Jefferson Lab, Newport News, VA 23606, USA}
\title{Spin-orbit correlations in the nucleon in the large-$N_{c}$ limit}
\begin{abstract}
We study the twist-3 spin-orbit correlations of quarks described by the nucleon matrix elements of
the parity-odd rank-2 tensor QCD operator (the parity-odd partner of the QCD energy-momentum
tensor).  Our treatment is based on the effective dynamics emerging from the spontaneous breaking of
chiral symmetry and the mean-field picture of the nucleon in the large-$N_c$ limit. The twist-3 QCD
operators are converted to effective operators, in which the QCD interactions are replaced by
spin-flavor-dependent chiral interactions of the quarks with the pion field.  We compute the nucleon
matrix elements of the twist-3 effective operators and discuss the role of the chiral interactions
in the spin-orbit correlations. We derive the first-quantized representation in the mean-field
picture and develop a quantum-mechanical interpretation.  The chiral interactions give rise to new
spin-orbit couplings and qualitatively change the correlations compared to the quark model
picture. We also derive the twist-3 matrix elements in the topological soliton picture where the
quarks are integrated out (skyrmion). The methods used here can be extended to other QCD operators
describing higher-twist nucleon structure and generalized parton distributions.
\end{abstract}
\maketitle
\tableofcontents
\section{Introduction}
Angular momentum is an essential aspect of nucleon structure in QCD.  The spin carried by quarks and
gluons is measured by operators connected with the parton distributions probed in polarized
deep-inelastic scattering and other experiments \cite{Aidala:2012mv,Deur:2018roz}.  The orbital
angular momentum is measured by operators derived from the energy-momentum tensor (EMT) connected
with the generalized parton distributions (GPDs) sampled in high-momentum-transfer exclusive
processes \cite{Ji:1996ek,Goeke:2001tz,Diehl:2003ny,Belitsky:2005qn,Leader:2013jra}.  The structures
can be discussed either in the language of fields or in the dual language of particles and their
interactions quantized at fixed light-front time. Substantial progress has been made in
characterizing the spin and orbital angular momentum distributions using theoretical methods and
experimental data.

Spin-orbit correlations represent a next step in the study of the internal angular momentum
structure of the nucleon in QCD \cite{Lorce:2014mxa}. They are measured by a certain twist-3
operator that can be viewed as the parity-odd analogue of the EMT. Similar to the EMT, the matrix
elements of this operator can be interpreted in terms of mechanical angular momentum and connected
with the measurable GPDs \cite{Ji:1996ek}. In the first-quantized particle picture, the matrix
elements of the quark spin-orbit correlation operator can be represented as
$\langle L_{z} S_{z} \rangle$, where $L_{z}$ and $S_{z}$ are the mechanical orbital and spin angular
momenta along the quantization axis. Using the equations of motion of the QCD fields, the twist-3
spin-orbit correlation operator can be related to the twist-2 vector current operator. This allows
one to express the quark spin-orbit correlations in terms of the well-known vector form factors of
the nucleon and a moment of the axial-vector GPDs \cite{Lorce:2014mxa}.

Nucleon structure at the hadronic scale is usually described in terms of effective degrees of
freedom evolving according to an effective dynamics. The spin structure has extensively been
discussed in effective models such as the quark model, chiral soliton model, and other
approaches. The spin-orbit correlations can be analyzed in the same effective models. However, in
this case a problem arises from the use of the QCD equations of motion in relating the twist-3 and 2
operators. The model estimates depend on which form of the QCD operator is naively identified with
the effective degrees of freedom, leading to inconsistent results. (The same problem arises in the
decomposition of the angular momentum density into spin and orbital angular momentum densities.) To
solve this problem one needs know how the effective degrees of freedom are related to QCD gauge
fields, and how the QCD operators are expressed in terms of the effective degrees of freedom.

Chiral symmetry breaking plays an essential role in the emergence of effective dynamics from QCD.
It accompanies the creation of mass from chromodynamic fields, causes the appearance of almost
massless hadronic excitations -- the pions, and determines the general structure of the
long-distance dynamics.  Chiral symmetry breaking is understood to be caused by topological vacuum
fluctuations of the QCD gauge fields, which induce zero modes of the fermion fields of definite
chirality; see Refs.~\cite{Diakonov:2002fq,Schafer:1996wv} for a review. An explicit realization of
this picture is the instanton vacuum, where the nonperturbative gauge fields are described
semiclassically as a superposition of instantons -- localized classical solutions of the Euclidean
Yang-Mills equations with topological charge $\pm 1$
\cite{Shuryak:1981ff,Shuryak:1982dp,Diakonov:1983hh,Diakonov:1985eg}.  The instantons have average
size $\bar\rho \sim$ 0.3 fm and average 4-dimensional separation $\bar R \sim$ 1 fm, and the system
is characterized by a small parameter in the form of the packing fraction
$\kappa \equiv \pi^2 \bar\rho^4 / \bar R^4 \approx 0.1$ (diluteness).  The instanton vacuum permits
systematic studies of chiral symmetry breaking and its consequences.  The effective dynamics is
obtained in the form of quarks with a dynamical mass $M \sim$ 0.3--0.4 GeV and spin-flavor dependent
interactions. It can be constructed and solved systematically using the $1/N_c$ expansion
(bosonization, saddle-point approximation)
\cite{Diakonov:1985eg,Diakonov:1986aj,Diakonov:1995qy,Kacir:1996qn}. The nucleon appears as a
self-consistent mean-field solution with a classical pion field (soliton) and quark moving in
single-particle orbits \cite{Diakonov:1987ty}.  The mean-field solution provides a specific
realization of the general picture of baryons in large-$N_c$ QCD \cite{Witten:1979kh}.  It
interpolates between quark model and chiral soliton and contains both as limiting cases
\cite{Praszalowicz:1995vi}.  It gives rise to a successful phenomenology and has been applied to a
range of hadronic observables and partonic structure; see Ref.~\cite{Christov:1995vm} for a review.

Recent work has studied the hadronic matrix elements of twist-3 QCD operators in the instanton
vacuum \cite{Kim:2023pll}.  In this approach it is possible to convert the QCD operators into
effective operators in the effective degrees of freedom, using same systematic approximations as in
derivation of effective dynamics \cite{Diakonov:1995qy,Balla:1997hf}.  The instantons provide an
explicit model of the nonperturbative gauge fields, which contribute to correlation functions of
gluonic operators and cause dynamical effects at the level of the effective theory.  It was found in
Ref.~\cite{Kim:2023pll} that the twist-3 spin-orbit correlation operators receive large
contributions from the instanton gauge fields. The instantons induce ``potential'' terms in the
effective operators, which complement the ``kinetic'' terms in the quark field
derivatives.\footnote{We use the terms ``kinetic'' and ``potential'' in the specific context of the
evaluation of QCD operators in the instanton vacuum, as introduced in Ref.~\cite{Kim:2023pll}. They
refer to the parts of the effective operators resulting from the quark field derivative $i\partial$
and the QCD gauge potential $A$ in QCD operators of the form $\bar\psi ... (i\partial - A)\psi$,
when these operators are evaluated in the instanton field.  We note that in the literature on the
gauge theories the derivative $i\partial$ is often referred to as the ``canonical'' momentum, and
the entire covariant derivative $(i\partial - A)$ is referred to as the ``kinetic'' momentum.}  The
potential terms describe spin-flavor dependent chiral interactions of the massive quarks with the
pion fields appearing in the bosonization of the effective dynamics. As a result, the effective
operators obey the same equation-of-motion relations in the effective theory as the original QCD
operators in QCD. This ensures that the results are independent of the form of the operator used
(``before'' or ``after'' the equations of motion) and enables consistent calculations of the matrix
elements in the effective theory.

In this work we study the twist-3 spin-orbit correlations in the nucleon on the basis of the
effective dynamics emerging from the spontaneous breaking of chiral symmetry and the mean-field
picture of the nucleon in the large-$N_c$ limit. The objectives are:

(i)~To derive the $N_c$-scaling of the spin-orbit correlations and their flavor components
(isoscalar/isovector). These scaling properties are general and independent of the dynamics, and can
be derived easily from the mean-field picture of the nucleon.

(ii)~To deliver a consistent numerical estimate of the spin-orbit correlations in nucleon, and to
quantify the effect of the instanton-induced chiral interactions on the spin-orbit correlations.
This is achieved by computing separately the contributions of the kinetic and potential terms of the
matrix elements of the effective operators. These results are specific to the instanton vacuum
dynamics and demonstrate the role of the nonperturbative gauge fields in the spin-orbit
correlations.

(iii)~To develop the mechanical interpretation of the spin-orbit correlations in the mean-field
picture of the nucleon in the large-$N_c$ limit (independent-particle motion), and to study the role
of spin-flavor interactions and relativistic motion.  The mean field picture enables a
first-quantized representation of the matrix elements of QCD operators and permits a simple
interpretation. It contains the quark model and chiral soliton as limiting cases and explains the
findings of these models.  These results depend mostly on general properties of effective dynamics,
not on the specifics of the instanton vacuum.

The plan of the article is as follows. In Sec.~\ref{sec:qcd} we discuss the QCD operators measuring
spin-orbit correlations and the form of the nucleon matrix elements.  In Sec.~\ref{sec:effective} we
review the effective theory emerging from chiral symmetry breaking in the instanton vacuum, and the
effective operators representing the twist-3 QCD operators and their properties In
Sec.~\ref{sec:nucleon} we summarize the description of the nucleon as a mean-field solution of the
effective theory and the calculation of matrix elements.  In Sec.~\ref{sec:spinorbit} we study the
spin-orbit correlations in the nucleon in the effective theory.  We first derive the general
$N_c$-scaling of the isospin components of the spin-orbit correlations.  We then compute the nucleon
matrix elements of the effective twist-3 operator, including the kinetic and potential terms, and
discuss their contributions.  In the large-$N_c$ mean-field picture the matrix element is obtained
in in first-quantized form, and we use this representation to explain the effect of chiral
interactions on the spin-orbit correlations.  We compute the contribution of the discrete level in
the quark single-particle spectrum and discuss the role of relativistic effects.  We derive the
gradient expansion of the matrix element and connect it with the topological soliton picture. We
then present the numerical results for the spin-orbit correlations.  In Sec.~\ref{sec:discussion} we
discuss further aspects of the twist-3 matrix elements and the spin-orbit correlations. We study the
``quark model'' and ``skyrmion'' limits of the large-$N_c$ mean-field picture of the spin-orbit
correlations. We compute the $t$-dependent form factors of the parity-odd tensor operator. We also
present the large-$N_c$ predictions for the $N$-$\Delta$ transition matrix elements.  In
Sec.~\ref{sec:summary} we summarize the results and discuss possible extensions to other structures.
\section{Spin-orbit correlations in QCD}
\label{sec:qcd}
\subsection{Operator and interpretation}
For the study of spin-orbit correlations of quarks in QCD one considers the parity-odd rank-2 tensor
operator \cite{Lorce:2014mxa}
\begin{align}
T_{5}^{\mu\nu}(x)
\equiv \bar{\psi}(x) i\gamma^{\mu} \gamma_{5} \overleftrightarrow{\nabla}^{\nu} \tau \psi (x),  
\label{operator_def}
\end{align}
where $\psi(x)$ and $\bar\psi(x)$ are the quark fields and
\begin{align}
\overleftrightarrow{\nabla}^\nu
&\equiv
\frac{1}{2} (\overrightarrow{\partial}^\nu - \overleftarrow{\partial}^\nu ) - i A^{\nu}(x)
\end{align}
is the QCD covariant derivative with gauge potential $A^{\nu}(x)$ in the fundamental representation
of the color group. Summation over color, spinor and flavor indices is implied in
Eq.~(\ref{operator_def}).  We assume two light quark flavors $\psi \equiv (\psi_u, \psi_d)$, and
$\tau$ denotes a $2\times 2$ flavor matrix.  In the theoretical analysis we consider the
flavor-singlet and -nonsinglet (isoscalar and isovector) operators
\begin{subequations}
\label{operator_isospin}
\begin{align}
(T_{5}^S)^{\mu\nu} &\equiv  T_{5}^{\mu\nu} [\tau = \bm{1}] = (T_{5}^{u + d})^{\mu\nu},
\\
(T_{5}^V)^{\mu\nu} &\equiv  T_{5}^{\mu\nu} [\tau = \tau^3] = (T_{5}^{u - d})^{\mu\nu},
\end{align}
\end{subequations}
where $\tau^3$ is the Pauli matrix.  Our definition of the operator Eq.~(\ref{operator_def}) is such
that the flavor-singlet operator coincides with the operator of Ref.~\cite{Lorce:2014mxa},
\begin{align}
(T_{5}^S)^{\mu\nu} &=  T_{q5}^{\mu\nu} (\text{Ref.~\cite{Lorce:2014mxa}}),
\end{align}
when accounting for differences in conventions (see details below).  The operator
Eq.~(\ref{operator_def}) is the parity-odd analogue of the quark contribution to the QCD EMT. Note
that the parity-odd operator is not a conserved current related to a global symmetry.

The tensor operator defined by Eq.~(\ref{operator_def}) has no definite symmetry and contains parts
of different spin and twist (see below). It obeys relations following from the QCD equations of
motion for the quark fields,
\begin{align}
(i\overrightarrow{\slashed{\nabla}} - m) \psi (x) = 0,
\hspace{2em}
\bar\psi (x) (-i\overleftarrow{\slashed{\nabla}} - m) = 0,
\end{align}
where $m$ is the matrix of the current quark masses.  The trace of the tensor operator is zero,
\begin{align}
(T_{5})^{\mu}_{\;\;\mu}(x) = 0 .
\label{qcdeom_trace}
\end{align}
The antisymmetric part of the tensor operator satisfies
\begin{align}
\partial_\mu T_{5}^{[\mu \nu]}(x) = \mathcal{O}(m) ,
\label{qcdeom_antisymmetric}
\end{align}
where $\partial_\mu$ is the total derivative (with respect to the entire $x$-dependence of the local
operator), and the right-hand is an operator proportional to the current quark masses. This relation
is analogous to the conservation law of the EMT; however, no corresponding relation holds for the
symmetric part.

The transition matrix element of the operator Eq.~(\ref{operator_def}) is taken between nucleon
states $| N \rangle \equiv |N (p, S_3)\rangle$ and $| N' \rangle \equiv |N (p', S_3')\rangle$ with
4-momenta $p$ and $p'$ and spin projections $S_3$ and $S_3'$, normalized to
$\langle N' | N \rangle = 2p^{0} (2\pi)^3 \delta^{(3)}(\bm{p}'-\bm{p}) \delta_{S'_{3}S_{3}}$.  $S_3$
and $S_3'$ denote the spin projections in the rest frame; the definition of the spin states for the
moving nucleons is discussed below.  The matrix element is decomposed in independent covariant
structures as follows \cite{Lorce:2014mxa}:
\begin{align}
\langle N' |T_{5}^{\mu\nu}(0) | N \rangle
&= \bar{u}' \left[
  \frac{P^{\{\mu}\gamma^{\nu\}}\gamma_{5}}{2} \tilde{A}(t)
+ \frac{P^{\{\mu}\Delta^{\nu\}}\gamma_{5}}{4M_{N}} \tilde{B}(t)
\right.
\nonumber \\
&+ \frac{P^{[\mu}\gamma^{\nu]}\gamma_{5}}{2} \tilde{C}(t)
 + \frac{P^{[\mu}\Delta^{\nu]}\gamma_{5}}{4M_{N}} \tilde{D}(t)
\nonumber \\
& \left. + M_{N} i \sigma^{\mu\nu} \gamma_{5} \tilde{F}(t) \right] u .
\label{parametrization}
\end{align}
Here and in the following we use the short-hand notation
\begin{align}
a^{\{\mu \nu \}} \equiv a^{\mu\nu} + a^{\nu\mu},
\hspace{2em}
a^{[\mu \nu]} \equiv a^{\mu\nu} - a^{\nu\mu}.
\label{sym_antisym}
\end{align}
$u \equiv u(p,S_{3})$ and $\bar u' \equiv \bar u(p',S_{3}')$ are the Dirac spinors with
normalization $\bar{u} u = 2 M_{N}$, where $M_{N}$ is the nucleon mass. The average and difference
of the initial and final 4-momenta are denoted as
\begin{align}
P \equiv (p'+p)/2,  \hspace{2em} \Delta \equiv p'-p,
\label{average_difference_momenta}
\end{align}
and the invariant momentum transfer is $t\equiv \Delta^{2}$.  The functions $\tilde A(t)$--$\tilde
F(t)$ are invariant form factors.  The parametrization Eq.~(\ref{parametrization}) takes into
account the tracelessness condition Eq.~(\ref{qcdeom_trace}). Unlike the parity-even EMT, there is
no constraint from current conservation on the matrix element of the parity-odd tensor operator, so
that it is parametrized by five instead of four independent form factors.
Equation~(\ref{parametrization}) is valid for any choice of nucleon spin states; the information on
the spin states is contained in the specific form of spinors. The invariant form factors can be
extracted from the operator matrix element with any choice of spin states.
Equation~(\ref{parametrization}) applies to both flavor-singlet and nonsinglet (isoscalar and
isovector) operators, Eq.~(\ref{operator_isospin}); the flavor (isospin) structure of the form
factors is denoted by superscripts $\tilde A^{u + d}, \tilde A^{u - d}$ etc.\ in the following.

The interpretation of the operator Eq.~(\ref{operator_def}) is developed in the context of
light-front quantization, where the quark-gluon structure of hadrons can be described in a particle
picture; see Ref.~\cite{Lorce:2014mxa} for details. Light-front 4-vector components are introduced
as $v^{\pm} \equiv (v^{0}\pm v^{z})/\sqrt{2}$ and $\bm{v}_T \equiv (v^x, v^y)$.  Particle number
operators are defined at fixed light-front time $x^+$. The second-quantized operator
\begin{align}
S_z \equiv \frac{1}{2} \int dx^- d^2 x_T \; \bar\psi (x) \gamma^+ \gamma_5 \psi (x)
\end{align}
measures the difference of the number of quarks with positive and negative spin projection along the
$z$-axis.  The operator
\begin{align}
L_z \equiv \int dx^- d^2 x_T \; \bar\psi (x) \gamma^+ (\bm{x}_T \times
i \overleftrightarrow{\bm{\nabla}})_z \psi (x)
\end{align}
measures the orbital angular momentum projection on the $z$-axis regardless of the spin; see
Refs.~\cite{Leader:2013jra,Lorce:2017wkb} for the derivation and connection with other definitions
of angular momentum in QCD. The operator
\begin{align}
C_{z} \equiv \int dx^- d^2 x_T \; \bar\psi(x) \gamma^+ \gamma_5 (\bm{x}_T \times
i \overleftrightarrow{\bm\nabla})_z \psi (x)
\label{spinorbit_operator}
\end{align}
measures the difference of the orbital angular momentum projections of quarks with positive and
negative spin projection, and thus describes the ``spin-orbit correlations'' of quarks in
light-front quantization. The operator in Eq.~(\ref{spinorbit_operator}) can be obtained from the
light-front components of the general tensor operator Eq.~(\ref{operator_def}) in a frame where the
momentum transfer is in the transverse direction,
$\Delta^+ = \Delta^- = 0, \bm{\Delta}_T \neq 0, \bm{\Delta}_T^2 = -t$.  The nucleon spin states are
chosen as light-front helicity states (spin states obtained from the rest-frame spin states by a
sequence of longitudinal and transverse light-front boosts).  The matrix element can then be
expressed in terms of the invariant form factors of Eq.~(\ref{parametrization}). The forward matrix
element, denoted by $C_z$, is obtained as \cite{Lorce:2014mxa}
\begin{align}
C_z &=   \left( -i \frac{\partial}{\partial \bm{\Delta}_{T}}
\times \frac{\langle N' | \bm{T}^{+T}_{5} | N \rangle}{2P^{+}} \right)_{z}
\bigg{|}_{\bm{\Delta}_T = 0}
\nonumber \\
&= \frac{1}{2} \left[ \tilde{A}(0) + \tilde{C}(0) \right].
\label{spinorbit_def}
\end{align}
Flavor (isospin) components of $C_z$ are defined as in Eq.~(\ref{operator_isospin}).

It is worth emphasizing that the light-front representation is needed only for the interpretation of
the spin-orbit correlations, not for their evaluation. The matrix element $C_z$ is expressed in
terms of the invariant form factors of the tensor operator, which can be evaluated using
formulations and methods unrelated to light-front quantization, e.g. Euclidean correlation
functions, or the $1/N_c$ expansion of the nucleon matrix elements in a 3-dimensional rotationally
invariant formulation \cite{Kim:2023xvw}.
\subsection{Relations for form factors}
The invariant form factors appearing in Eq.~(\ref{spinorbit_def}) can be related to nucleon matrix
elements of other known QCD operators. The form factor $\tilde{A}(t)$ can be obtained as the second
moment of the twist-2 axial-vector GPD \cite{Lorce:2014mxa}
\begin{align}
\tilde{A}(t) = \int^{1}_{-1} dx\, x \tilde{H}(x,\xi,t).
\label{Atilde_from_gpd}
\end{align}
Here $x$ is the longitudinal momentum fraction carried by the quark, $\xi \equiv -\Delta^{+}/2P^{+}$
is the skewness variable, and the GPD is defined as
\begin{align}
&\int \frac{d z^{-}}{2\pi} e^{i x P^{+} z^{-}}
\langle N' | \bar{\psi}(-z/2) \gamma^{+} \gamma^{5} \psi(z/2) | N \rangle_{z^+, \bm{z}_T = 0}
\nonumber \\
&= \bar{u}' \left[ \frac{\gamma^{+}\gamma^{5}}{P^{+}} \tilde{H}(x,\xi,t) + \ldots \right] u,
\end{align}
where the ellipsis represents structures that are not relevant to the current study.

Furthermore, using the QCD equations of motion, the antisymmetric part of the tensor operator
Eq.~(\ref{operator_def}) can be expressed as the total derivative of the vector current operator,
\begin{align}
\frac{1}{2} T^{[\mu \nu] }_{5}(x) &=
\frac{1}{2}\bar{\psi} \gamma^{[\mu} \gamma_{5} i \overleftrightarrow{\nabla}^{\nu ]} \tau \psi
\nonumber \\
&= - \frac{1}{4} \epsilon^{\mu\nu\alpha\beta}
\partial_{\alpha}(\bar{\psi} \gamma_{\beta} \tau \psi ) + \mathcal{O}(m).
\label{eom_qcd}
\end{align}
The relation holds for any flavor matrix $\tau$; we have neglected terms proportional to the current
quark masses. In our convention $\epsilon^{0123}=1$ and
$\gamma_{5} = - i \gamma_{0}\gamma_{1}\gamma_{2}\gamma_{3}$ \cite{Berestetskii:1982qgu}.
Equation~(\ref{eom_qcd}) leads directly to Eq.~(\ref{qcdeom_antisymmetric}).
Equation~(\ref{eom_qcd}) makes it possible to express the form factors $\tilde{C}(t)$ and
$\tilde{F}(t)$ in Eq.~(\ref{parametrization}) in terms of the well-known form factors of the vector
current operator \cite{Lorce:2014mxa}:
\begin{subequations}
\label{formfactors_from_vector}
\begin{align}
\tilde{C}(t) &= - F_{1}(t),
\\
\tilde{F}(t) &= - \frac{1}{2} G_{E}(t)
= - \frac{1}{2} \left[ F_{1}(t) + \frac{t}{4M^{2}_N} F_{2}(t) \right],
\end{align}
\end{subequations}
where $F_{1, 2}(t)$ are the Dirac/Pauli form factors and the combination $G_{E}(t)$ is the Sachs
electric form factor. The form factors here correspond to the vector currents with flavor structure
$u + d$ or $u - d$ ($\tau = \bm{1}$ or $\tau^3$), Eq.~(\ref{eom_qcd}), and do not contain any quark
charge factors.  In particular, the values of the form factors at zero momentum transfer are given
by the number of valence quarks; for the proton state
\begin{subequations}
\begin{align}
F_{1}^{u + d}(0) &= G_{E}^{u + d}(0) = 3 ,
\\
F_{1}^{u - d}(0) &= G_{E}^{u - d}(0) = 1 .
\end{align}
\end{subequations}
Combining Eqs.~(\ref{spinorbit_def}), (\ref{Atilde_from_gpd}) and (\ref{formfactors_from_vector}),
it is possible to express the spin-orbit correlation matrix element in terms of known quantities:
\begin{subequations}
\begin{align}
C^{u+d}_{z}= \frac{1}{2} \int^{1}_{-1} dx \, x \tilde{H}^{u+d} (x, 0, 0)
- \frac{1}{2} F^{u+d}_{1}(0),
\\
C^{u-d}_{z}= \frac{1}{2} \int^{1}_{-1} dx \, x \tilde{H}^{u-d} (x, 0, 0)
- \frac{1}{2} F^{u-d}_{1}(0).
\end{align}
\end{subequations}
\section{Effective dynamics and operators}
\label{sec:effective}
\subsection{Chiral symmetry breaking from instantons}
The mechanism of chiral symmetry breaking in QCD and the foundations of the instanton vacuum are
described in detail in Refs.~\cite{Diakonov:2002fq,Schafer:1996wv}; here we briefly summarize the
elements used in the present study. Chiral symmetry breaking in Euclidean QCD is caused by
topological fluctuations of the gauge fields.  They induce localized zero modes of the fermion
fields with definite chirality, which delocalize to form a chiral condensate. An explicit
realization of this picture is provided by instanton vacuum
\cite{Diakonov:2002fq,Schafer:1996wv}. The gauge fields are described as a superposition of
instantons and antiinstantons ($I$ and $\bar I$), classical topological fields representing
solutions of the Yang-Mills equations (tunneling trajectories between field configurations with
different winding number), surrounded by quantum fluctuations.  The $I$'s and $\bar I$'s are
parametrized by collective coordinates (position, color orientation, size), and the functional
integral over the gauge fields becomes a statistical mechanics in these variables.  The properties
of the system can be derived from empirical observations \cite{Shuryak:1981ff,Shuryak:1982dp} or
from theoretical calculations with instanton interactions \cite{Diakonov:1983hh,Shuryak:1988rf}, and
can be validated by lattice QCD simulations; see Ref.~\cite{Diakonov:2002fq,Schafer:1996wv} for a
review. The instantons have average size $\bar\rho \sim$ 0.3 fm and average 4-dimensional separation
$\bar R \sim$ 1 fm.  As a result, the system is characterized by a small parameter in the form of
the packing fraction $\kappa \equiv \pi^2 \bar\rho^4 / \bar R^4 \approx 0.1$ (diluteness). This
parameter enables systematic studies of chiral symmetry breaking and hadronic correlation functions.

Chiral symmetry breaking in the instanton vacuum can be studied in the $1/N_c$ expansion
\cite{Diakonov:1985eg,Diakonov:1986aj,Diakonov:1995qy,Kacir:1996qn}.  The instantons couple to light
fermions through the localized zero modes of the Dirac operator of definite chirality. Each
($I\bar{I}$) induces a multifermion vertex with spin-flavor structure of the form
\begin{align}
\propto \mathrm{det}_{f'f} \; \bar\psi_{f'}(z) \frac{1\pm\gamma_{5}}{2} \psi_{f}(z),
\label{det}
\end{align}
where $z$ is the $I(\bar{I})$ position; the color orientation is integrated over.  The interaction
actually occurs over a finite range of Euclidean distances $\sim \bar\rho$ around $z$, determined by
the extent of the zero mode wave function; the spacetime structure is not indicated in
Eq.~(\ref{det}); see Ref.~\cite{Diakonov:1995qy} for details. In the large-$N_c$ limit, the ground
state of the interacting fermion system can be obtained by applying the saddle point approximation
to the functional integral.  A dynamical quark mass $M > 0$ appears due to the finite density of
instantons. Parametrically it is of the order
\begin{align}
M^2 \sim \kappa \bar\rho^{-2},
\end{align}
which means that it is ``parametrically small'' compared to the inverse instanton size.  Numerically
it takes values $M \approx$ 0.3--0.4 GeV, which are those of a typical constituent quark mass.
Together with the dynamical quark mass a vacuum condensate of the quark fields arises, representing
the order parameter of chiral symmetry breaking.

In the large-$N_c$ limit the system of interacting quarks can be bosonized by introducing local
meson fields representing the fermion bilinears
\begin{align}
\bar\psi_{f'}(x) \frac{1\pm\gamma_{5}}{2} \psi_{f}(x).
\end{align}
For describing the effective dynamics at momenta $p^2 \sim M^2$ it is sufficient to restrict the
meson fields to the ``chiral phase'' parametrizing chiral rotations of the vacuum condensate
(Nambu-Goldstone boson modes).  It is defined as
\begin{align}
U^{\gamma_{5}}(x) &\equiv \frac{1+\gamma_{5}}{2} U(x) + \frac{1-\gamma_{5}}{2} U^{\dagger}(x),
\label{U_gamma5_def}
\end{align}
where
\begin{align}
U(x) = \exp[i \pi^{a}(x) \tau^{a}/F_{\pi}]
\end{align}
is a $2 \times 2$ unitary matrix field, $\pi^a(x) (a = 1,2,3)$ is the pion field with physical
normalization, and $F_\pi$ is the pion decay constant with the empirical value 93 MeV.  The
effective dynamics emerging from chiral symmetry breaking can then be represented in so-called
semi-bosonized form, as a functional integral over the massive quark fields and the chiral meson
field\footnote{The effective dynamics arising from chiral symmetry breaking in the instanton vacuum
is originally obtained as a Euclidean functional integral (imaginary time)
\cite{Diakonov:1995qy}. We present here the form after analytic continuation to Minkowski space
(real time), which can be directly connected with the equation of motion of the fields and the form
of the effective operators.
\label{footnote:minkowski}}
\begin{subequations}
\label{effective_theory}
\begin{align}
& Z_{\mathrm{eff}} \equiv \int \mathcal{D} U \!\int\mathcal{D}\bar\psi \mathcal{D}\psi \;
\exp(i S_{\mathrm{eff}}[\bar\psi, \psi, U]),
\\[1ex]
& S_{\mathrm{eff}}[\bar\psi, \psi, U] \equiv \int d^{4} x \, \bar\psi (x)
[ i\slashed{\partial} - M U^{\gamma_{5}}(x) ] \psi (x).
\end{align}
\end{subequations}
It describes the dynamics of the massive quark field modes with momenta
$p^2 \sim M^2 \ll \bar\rho^{-2}$.  The instanton size $\bar\rho^{-2}$ acts as the ultraviolet cutoff
of the effective theory.  In the actual Euclidean effective theory the cutoff is implemented by the
finite extent of the zero mode wave function, which provides a finite range of the multifermion
interaction and a momentum dependence of the dynamical quark mass. For describing the dynamics at
momenta $p^2 \sim M^2 \ll \bar\rho^{-2}$ these details are not needed, and the cutoff can be
implemented through a generic ultraviolet regularization (see Secs.~\ref{sec:nucleon} and
\ref{sec:spinorbit}).  In the effective theory described by Eq.~(\ref{effective_theory}) the massive
quark/antiquark fields is coupled to the chiral meson field, which is itself a composite of massive
quark and antiquark fields. The form of the coupling is dictated by chiral invariance and can be
inferred from general considerations \cite{Diakonov:1983bny}; as such it is more general than the
specific dynamics of the instanton vacuum. The strength of the coupling is determined by the
dynamical quark mass $M$. The quark-pion coupling constant in conventional terms (with the physical
normalization of the pion field) is given by $M/F_\pi \sim 4$; the system is thus strongly coupled
and needs to be solved using nonperturbative methods.

Hadronic correlation functions in the effective theory can be computed in the $1/N_c$ expansion
\cite{Diakonov:1985eg,Diakonov:1986aj,Kacir:1996qn}.  It enables a semiclassical expansion of the
functional integral over the chiral meson field in Eq.~(\ref{effective_theory}) (saddle point
approximation).  In mesonic correlation functions the saddle point is the trivial classical field
$U_{\rm cl}(x) \equiv 1$, and the integral becomes a Gaussian integral over the quantum fluctuations
around this background.  The two-point function in the pseudoscalar isovector channel exhibits the
pion pole with mass $M_\pi^2$ proportional to the current quark mass. The constant $F_\pi^2$ is
extracted from the field normalization and obtained as a loop integral of the massive quark
propagator; it is parametrically of the order $F_\pi^2 \sim N_c M^2 \log (M\bar\rho)$ and
numerically in good agreement with the empirical value. Higher-point functions describe pion-current
and pion-pion interactions The two-point function in the pseudoscalar isoscalar channel exhibits a
massive $\eta'$ pole with $M_{\eta'}^2 \sim \bar\rho^{-2}$. The scalar isoscalar channel also
exhibits a massive pole; for other meson channels (vector, tensor) see Ref.~\cite{Schafer:1996wv}.

In baryonic correlation functions the saddle point is characterized by a nontrivial classical field
$U_{\rm cl}(x) \neq 1$ with a distinctive spatial and flavor dependence (``soliton'')
\cite{Diakonov:1987ty}.  The resulting mean-field picture of baryons gives rise to a rich structure
and is discussed in Sec.~\ref{sec:nucleon}.
\subsection{Effective operators from instantons}
\label{subsec:effective_operators}
In the effective theory defined by Eq.~(\ref{effective_theory}), QCD operators are represented by
effective operators in the effective degrees of freedom.  In the context of the instanton vacuum,
the effective operators can be derived from the QCD operators using the same approximations as in
the derivation of effective action (packing fraction expansion, $1/N_c$ expansion, bosonization)
\cite{Diakonov:1995qy,Balla:1997hf,Kim:2023pll}.  It is assumed that the QCD operators are
normalized at the scale $\mu = \bar\rho^{-1}$, which is the cutoff scale of the effective theory.
QCD operators involving the gluon field are evaluated in the classical gluon field of the
instantons, which couples them to the quarks through the zero modes and converts them into
multi-fermion operators.  The multi-fermion operators in turn can be bosonized in the $1/N_c$
expansion and become operators in the massive quark fields coupled to the chiral meson field.  These
effective operators can then be inserted in correlation functions in the effective theory.  The
possibility of a systematic derivation of the effective operators represents a major advantage of
instanton approach and enables studies of the hadronic matrix elements of a wide range of QCD
operators.

The effective operator representing the parity-odd tensor QCD operator Eq.~(\ref{operator_def}) was
derived in Ref.~\cite{Kim:2023pll}. Through a series of steps (bosonization, partial contraction of
the quark fields) the effective multi-fermion operator induced by the instantons was converted to a
two-fermion operator in the background of the chiral meson field. It is given by\footnote{For
simplicity we use the same notation for the effective operator as for the original QCD operator. The
two versions can be distinguished by the context.}
\begin{align}
&T_{5}^{ \mu\nu }(x) \equiv  \bar{\psi}(x) \left\{ i\gamma^{ 
\mu} \gamma_{5}    \overleftrightarrow{\partial}^{\nu } \tau  \right.
\nonumber \\
& \left. + \frac{M}{4} \left[  \gamma^{\mu} \gamma^{\nu} \gamma_{5} \tau U^{\gamma_{5}}(x)
- U^{\gamma_{5}}(x)\tau \gamma^{\nu} \gamma^{\mu} \gamma_{5} \right] \right\} \psi(x) ,
\label{effective_operator_def}
\end{align}
where $\tau$ is the flavor matrix $(\tau = 1$ or $\tau^3$).  This form of the effective operator
covers the spin-2 (symmetric traceless tensor) and spin-1 (antisymmetric tensor) projections; in the
spin-0 projection (trace) additional terms appear, which are not included in
Eq.~(\ref{effective_operator_def}).  The first term in Eq.~(\ref{effective_operator_def}) is
referred to as the kinetic term. It arises from the quark field derivative in the covariant
derivative of the QCD operator.  The second term in Eq.~(\ref{effective_operator_def}) is referred
to as potential term. It arises from the gauge potential in the covariant derivative of the QCD
operator, which is evaluated in the instanton field and results in a coupling of the quark to the
chiral fields, proportional to the dynamical quark mass $M$. This term represents dynamical effect
of instantons, and its form is conditioned by the specific spin-flavor structure of the instanton
field. It affects the various spin projections of the effective operator in different ways.

In the twist-2 projection of the effective operator the potential term is absent. This can be seen
immediately from the fact that the structure involves $\gamma^\mu \gamma^\nu$ and $\gamma^\nu
\gamma^\mu$, which gives zero when forming a symmetric traceless tensor. The twist-2 effective
operator is given entirely by the kinetic term [see Eq.~(\ref{sym_antisym})] \begin{align}
&\frac{1}{2} T_{5}^{ \{\mu\nu \}}(x) - \mathrm{trace}
\nonumber \\
&= \frac{1}{2} T_{5}^{ \{\mu\nu \}}(x)[\textrm{kin}]  - \mathrm{trace}
\nonumber \\
&= \frac{1}{2} \bar{\psi}(x)  i\gamma^{ 
\{\mu} \gamma_{5}    \overleftrightarrow{\partial}^{\nu\} } \tau \psi(x) - \mathrm{trace}.
\end{align}
In the twist-2 case, the instanton vacuum justifies the use of a naive definition of the effective
operator.

In the twist-3 projection of the effective operator the potential term is present
\cite{Kim:2023pll}.  Antisymmetrizing the terms in Eq.~(\ref{effective_operator_def}) one obtains
[see Eq.~(\ref{sym_antisym})]
\begin{subequations}
\label{effective_operator_twist3}
\begin{align}
& \frac{1}{2} T_{5}^{ [\mu\nu ]}(x) =
\frac{1}{2} T_{5}^{ [\mu\nu ]}(x)[\textrm{kin}] +
\frac{1}{2} T_{5}^{ [\mu\nu ]}(x)[\textrm{pot}],
\\[2ex]
& \frac{1}{2} T_{5}^{ [\mu\nu ]}(x)[\textrm{kin}]
= \frac{1}{2} \bar{\psi}(x) i\gamma^{ [\mu} \gamma_{5}
\overleftrightarrow{\partial}^{\nu]} \tau \psi(x),
\label{antisymmetric_kin}
\\[1ex]
& \frac{1}{2} T_{5}^{ [\mu\nu] }(x)[\textrm{pot}]
= -\frac{M}{4} \bar{\psi}(x) i\sigma^{ \mu \nu} \gamma_{5} \{ \tau, U^{\gamma_{5}}(x) \} \psi(x),
\label{antisymmetric_pot}
\end{align}
\end{subequations}
where $\{ \tau, U^{\gamma_5} \} \equiv \tau U^{\gamma_5} + U^{\gamma_5} \tau$.  The potential term
has important dynamical effects. It ensures that the effective operator obeys the same
equation-of-motion relation Eq.~(\ref{eom_qcd}) as the original QCD operator.  Using the equation of
motion of the quark fields in the effective theory,
\begin{subequations}
\label{eom_effective}
\begin{align}
[ i \overrightarrow{\slashed{\partial}} - M U^{\gamma_5}(x)] \psi (x)
&= 0,
\\
\bar\psi (x) [ -i \overleftarrow{\slashed{\partial}} - M U^{\gamma_5}(x)]
&= 0,
\end{align}
\end{subequations}
the effective operator Eq.~(\ref{effective_operator_twist3}) can be converted to the total
derivative of the vector current in the effective theory,
\begin{align}
& \frac{1}{2} T_{5}^{ [\mu\nu ]}(x) = - \frac{1}{4}
\epsilon^{\mu\nu\alpha\beta} \partial_\alpha \left[\bar\psi (x) \gamma_\beta \tau \psi (x) \right].
\label{total_derivative_effective}
\end{align}
This remarkable result attests to the consistency of the approximations in the derivation of the
effective dynamics and the effective operators from the instanton vacuum. The potential term also
qualitatively changes the spin-orbit correlations in the nucleon. This is demonstrated at the level
of nucleon matrix elements in Sec.~\ref{sec:nucleon}.
\subsection{Bosonized effective operators}
\label{subsec:bosonized}
The effective theory is originally obtained in the semi-bosonized form of
Eq.~(\ref{effective_theory}), as a functional integral over the massive quark and chiral meson
fields.  It can also be represented in fully bosonized form, by integrating out the quark fields in
the background of the chiral meson fields.  In the fully bosonized theory the effective operators
are represented by operators in the chiral meson fields.  This formulation is useful for making
connection with the Lagrangian of chiral effective field theory (ChEFT)
\cite{Weinberg:1968de,Gasser:1983yg}, and with the picture of the nucleon as a topological soliton
in the pion field in large-$N_c$ limit (``skyrmion'')
\cite{Skyrme:1961vq,Adkins:1983ya,Zahed:1986qz}.  Here we derive the fully bosonized form of the
effective operator representing the parity-odd tensor QCD operator Eq.~(\ref{operator_def}) and
discuss its properties.

Integrating over the quark fields in Eq.~(\ref{effective_theory}), the effective theory is obtained
in the form
\begin{subequations}
\label{effective_theory_bosonized}
\begin{align}
Z_{\mathrm{eff}} &= \int \mathcal{D} U \, \exp (i W_{\mathrm{eff}}[U]) ,
\\
W_{\mathrm{eff}}[U] &\equiv \mathrm{log \, Det}[i \slashed{\partial} - M U^{\gamma_{5}}] 
\end{align}
\end{subequations}
(the expressions are again presented for Minkowskian metric, see Footnote~\ref{footnote:minkowski}).
The effective action is given by the fermion determinant in the background of the chiral field.  Its
dependence on the chiral field can be made explicit by performing an expansion in derivatives of the
chiral field (gradient expansion); for practical techniques see e.g.\ Ref.~\cite{Diakonov:1987ty}.
One obtains
\begin{align}
W_{\mathrm{eff}}[U]
& = \int d^{4} x \left\{ \frac{F^{2}_{\pi}}{4}
\mathrm{tr}_{\textrm{fl}}[\partial^{\alpha} U^{\dagger} \partial_{\alpha} U] \right.
\nonumber \\
& \left. \phantom{\frac{0}{0}} + \textrm{terms}(\partial U^{4}) + \ldots \right\},
\label{chiral_lagrangian}
\end{align}
where the trace is over flavor indices.  The integrand in Eq.~(\ref{chiral_lagrangian}) is the
chiral Lagrangian. The structure of the terms is determined by chiral invariance. The leading term
is the unique $(\partial U)^2$ term; the $(\partial U)^4$ terms are given in
Ref.~\cite{Gasser:1983yg}. The constant $F^{2}_{\pi}$ and those of the higher-order terms are
obtained here as loop integrals of the massive quark propagator in the trivial background field
($U = 1$). As such they can be predicted in terms of the dynamical scales $M$ and $\bar\rho^{-1}$
(or the UV cutoff) and compared with the empirical values extracted from ChEFT calculations
\cite{Choi:2003cz,Goeke:2007bj}. The higher-order expansion of the fermion determinant also yields
the Wess-Zumino term of the chiral Lagrangian, which is connected with the topological properties of
ChEFT and enables the identification of soliton solutions as baryons \cite{Zahed:1986qz}.

In the bosonized effective theory Eq.~(\ref{effective_theory_bosonized}), the effective operators
are given by the fermionic average of the quark operators in the background of the classical chiral
field,
\begin{align}
O[\bar\psi, \psi, U] \; \rightarrow \;
\langle O[\bar\psi, \psi, U] \rangle_{U} \equiv O[U].
\end{align}
The average is computed in leading order of the $1/N_c$ expansion and given by loop integrals of the
quark propagator in the background of the chiral field,
\begin{subequations}
\begin{align}
&\textrm{T} \, \psi (x) \bar\psi (y) = i G(x, y | U),
\\
&[i \slashed{\partial}_x - M U^{\gamma_{5}}(x)] \, G(x, y | U) = \delta^{(4)} (x - y),
\end{align}
\end{subequations}
where $\textrm{T}$ denotes the time-ordered product.
The loop integrals can be expanded in gradients of the chiral field in the same manner as 
the effective action Eq.~(\ref{chiral_lagrangian}) \cite{Diakonov:1987ty}. For example, for the vector 
current operator appearing in Eq.~(\ref{eom_qcd}) (isoscalar or isovector)
\begin{align}
V^{\mu}(x) \equiv \bar\psi (x) \gamma^\mu \tau \psi(x) \hspace{2em} [\tau = \bm{1}, \tau^3]
\end{align}
the average in the background field is 
\begin{align}
\langle V^{\mu}(x)\rangle_U = i N_c \textrm{tr}[ G(x, y | U) \, \gamma^\mu \tau ],
\end{align}
where the trace runs over spinor and flavor indices. The gradient expansion in lowest nonzero order
gives
\begin{subequations}
\label{vector_current_gradient}
\begin{align}
&\langle V^{\mu}(x) \rangle_{U}
\nonumber
\\[1ex]
&\stackrel{(a)}{=} \frac{N_{c}}{24\pi^{2}} \epsilon^{\mu\alpha\beta\gamma}
\mathrm{tr}_{\textrm{fl}} [L_{\alpha} L_{\beta} L_{\gamma} ] \hspace{2em} [\tau=\bm{1} ] 
\\[1ex]
&\stackrel{(b)}{=} \frac{F^{2}_{\pi}}{2i}
\mathrm{tr}_{\textrm{fl}} [(L^{\mu} + R^{\mu})\tau^{3}] \hspace{2em} [\tau=\tau^{3} ]
\end{align}
\end{subequations}
(Here and in the following we present the expressions for $\tau = \bm{1}$ and $\tau^3$ in the same
formula and distinguish them by labels.) Here
\begin{subequations}
\label{left_right_currents}
\begin{align}
L^{\mu}(x) \equiv U^{\dagger}(x)\partial^{\mu}U(x), 
\\
R^{\mu}(x) \equiv U(x)\partial^{\mu}U^{\dagger}(x)
\end{align}
\end{subequations}
are the left and right currents of the chiral field.  The isoscalar current in
Eq.~(\ref{vector_current_gradient}) is of order $\partial U^{3}$; the isovector current is of order
$\partial U$. The isoscalar current obtained from the gradient expansion agrees with the well-known
expression obtained as the $U(1)$ symmetry current from the Wess-Zumino term of the chiral
Lagrangian \cite{Zahed:1986qz}.

We now derive the bosonized effective operators for the antisymmetric part of the tensor operator
Eq.~(\ref{effective_operator_def}) and verify the equation-of-motion relation for the bosonized
operators.  According to Eq.~(\ref{eom_qcd}) the bosonized operators must satisfy
\begin{subequations}
\begin{align}
&\frac{1}{2}
\langle T^{[\mu \nu]}_{5}(x) \rangle_{U} = - \frac{1}{4} \epsilon^{\mu \nu \alpha \beta }
\partial_{\alpha} \langle V_{\beta}(x) \rangle_{U}
\nonumber \\[1ex]
& \stackrel{(a)}{=} -\frac{N_{c}}{32\pi^{2}} \partial_{\alpha} \; 
\mathrm{tr}_{\textrm{fl}} [L^{\alpha} L^{[\mu} L^{\nu ]}]
\hspace{2em} [\tau=\bm{1} ] 
\\[1ex]
& \stackrel{(b)}{=} -\frac{F^{2}_{\pi}}{8i} \epsilon^{\mu \nu \alpha \beta} 
\partial_{\alpha} \; \mathrm{tr}_{\textrm{fl}} [(L_{\beta} + R_{\beta})\tau^{3}]
\hspace{2em} [\tau=\tau^{3} ]
\end{align}
\end{subequations}
where we have evaluated the R.H.S.\ of Eq.~(\ref{eom_qcd}) with the leading-order gradient expansion
results of Eq.~(\ref{vector_current_gradient}).  These relations are realized in a non-trivial way
in the effective theory.  From the kinetic term of the effective operator,
Eq.~(\ref{antisymmetric_kin}), we obtain
\begin{subequations}
\label{tensor_gradient_kin}
\begin{align}
&\frac{1}{2}\langle T^{[\mu \nu]}_{5}(x) \, [\textrm{kin}] \; \rangle_{U} 
\nonumber
\\[1ex]
& \stackrel{(a)}{=}
+ \frac{N_{c}}{48\pi^{2}}
\mathrm{tr}_{\textrm{fl}} [\partial_{\alpha} L^{[\mu} L^{\nu ]} L^{\alpha}] 
\hspace{2em} [\tau=\bm{1} ] 
\\[1ex]
& \stackrel{(b)}{=}
- \frac{F^{2}_{\pi}}{8i} \epsilon^{\mu \nu \alpha \beta} \partial_{\alpha} 
\mathrm{tr}_{\textrm{fl}} [(L_{\beta} + R_{\beta})\tau^{3}]  \hspace{2em} [\tau=\tau^{3} ]
\end{align}
\end{subequations}
(In expressions such as $\partial L L...L$ the derivative acts only on the nearest $L$ function;
cases where the derivative acts on multiple functions are indicated by parentheses.)  From the
potential term of the effective operator, Eq.~(\ref{antisymmetric_pot}), we obtain
\begin{subequations}
\label{tensor_gradient_pot}
\begin{align}
&\frac{1}{2}\langle T^{[\mu \nu]}_{5}(x) \, [\textrm{pot}] \; \rangle_{U}
\nonumber
\\[1ex]
& \stackrel{(a)}{=}
- \frac{N_{c}}{32\pi^{2}} \partial_{\alpha}
\mathrm{tr}_{\textrm{fl}} [L^{\alpha} L^{[\mu} L^{\nu ]} ]  
\nonumber
\\[1ex] 
& \hspace{1.3em} - \frac{N_{c}}{48\pi^{2}}
\mathrm{tr}_{\textrm{fl}} [\partial_{\alpha} L^{[\mu} L^{\nu ]} L^{\alpha} ] 
\hspace{2em} [\tau=\bm{1} ] 
\\[1ex]
& \stackrel{(b)}{=} \;
0 \hspace{2em} [\tau=\tau^{3} ]
\end{align}
\end{subequations}
One sees that the bosonized effective operators satisfy the equation-of-motion relation in both the
isoscalar and isovector channels.  In the isoscalar channel the kinetic and potential term are of
the same order, and both are required to obtain the correct result.  In the isovector channel the
kinetic term gives the entire result at the leading nonzero order in $\partial U$ (this was already
noted in Ref.~\cite{Kim:2023pll}); the potential term is zero at this order, producing the correct
total result.  This shows the role of the potential term of the effective operator in the spin-orbit
correlations.  The bosonized operators derived here are used in the gradient expansion of the
nucleon matrix elements in Sec.~\ref{subsec:gradient_expansion}.
\section{Nucleon in large-$N_c$ limit}
\label{sec:nucleon}
\subsection{Classical mean-field solution}
The nucleon appears as a self-consistent mean-field solution of the effective theory
Eq.~(\ref{effective_theory}) in large-$N_c$ limit \cite{Diakonov:1987ty}.  This represents a
realization of the general soliton picture of baryons in large-$N_c$ limit of QCD
\cite{Witten:1979kh} with the specific effective dynamics obtained by chiral symmetry breaking from
the instanton vacuum.  The mean-field solution is also known as chiral quark-soliton model
(especially when referring to it in a context more general than the instanton vacuum) and has been
used extensively in studies of hadronic properties \cite{Wakamatsu:1990ud,Christov:1995vm} and
parton distributions \cite{Diakonov:1996sr,Diakonov:1997vc}.  Here we briefly summarize the elements
used in present study of spin-orbit correlations.

%
%
\begin{figure}[t]
\includegraphics[width=0.95\columnwidth]{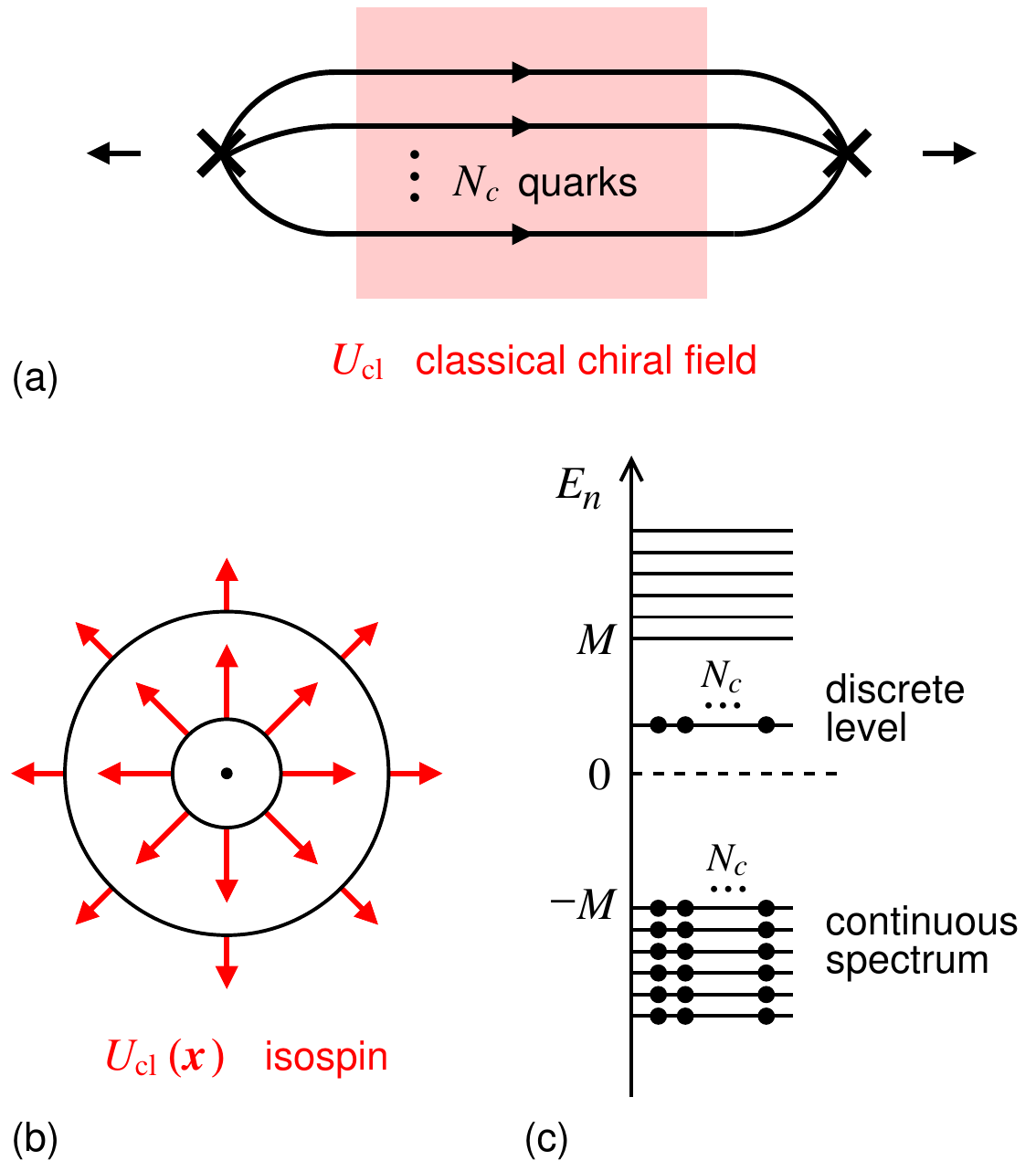}
\caption{Emergence of baryons from the effective dynamics. (a)~Correlation function of baryon
currents in large-$N_c$ limit. (b)~Spatial form of classical chiral field (hedgehog),
Eq.~(\ref{hedgehog}).  (c)~Quark single-particle spectrum in background of classical chiral field,
Eq.~(\ref{spectrum}).}
\label{fig:baryon}
\end{figure}
Correlation functions of baryon currents composed of $N_c$ quark fields (in a totally antisymmetric
color-singlet configuration) are computed in the effective theory Eq.~(\ref{effective_theory}) in
the $1/N_c$ expansion (saddle point approximation).  The saddle point solution is characterized by a
nontrivial classical chiral field (see Fig.~\ref{fig:baryon}a).  It is time-independent (static) at
finite Euclidean times and has a particular spatial form, in which the isospin direction is aligned
with the spatial direction (``hedgehog'', see Fig.~\ref{fig:baryon}b),
\begin{align}
U_{\rm cl}(\bm{x}) = \exp[i \bm{n} \cdot \bm{\tau} P(r)],
\label{hedgehog}
\end{align}
where $\bm{n} \equiv \bm{x}/|\bm{x}|$ is the radial unit vector and $P(r)$ is a radial profile
function with $P(0) = \pi$ and $P(\infty) = 0$ ($r \equiv |\bm{x}|$).  The field is invariant under
combined spatial and flavor rotations and encodes the emergent spin-flavor symmetry of baryons in
large-$N_c$ limit. The quark field degrees of freedom can be described in a first-quantized
picture. The quarks move in single-particle orbits in the classical chiral field, described by the
Hamiltonian (given here in Minkowskian metric)
\begin{subequations}
\label{hamiltonian}
\begin{align}
H \equiv -i \gamma^0 \gamma^k \partial_{k} + \gamma^0 M U^{\gamma_{5}}_{\rm cl} (\bm{x}),
\\[1ex]
U^{\gamma_5}_{\rm cl} (\bm{x}) = \exp[i \gamma_5 \bm{n} \cdot  \bm{\tau} P(r)] .
\end{align}
\end{subequations}
The single-particle wave functions $\Phi_{n}(\bm{x})$ and energy levels $E_{n}$ are obtained by
diagonalizing the Hamiltonian,
\begin{align}
H \Phi_{n} (\bm{x}) = E_{n} \Phi_{n} (\bm{x}).
\label{spectrum}
\end{align}
The spectrum includes a discrete level with energy $E_{\rm lev} < M$, and the continuous spectra of
levels with negative energies $E < -M$ and positive energies $E > M$ (see Fig.~\ref{fig:baryon}c).
In the ground state of a baryon with baryon number $B = 1$ the discrete level and the negative
continuum are occupied by $N_c$ quarks each.  The energy of the system in leading order of $1/N_c$
is given by the sum of the energies of the discrete level and the negative continuum, with the
vacuum energy subtracted (classical field $U_{\rm cl} = 1$, energy levels $E_n^{(0)}$)
\begin{align}
E[U_{\rm cl}] &= N_c E_{\rm lev} \; + \; N_c \! \sum_{n\; {\rm neg. cont.}} \! [E_n - E_n^{(0)}].
\label{energy_sum}
\end{align}
The actual classical field is determined by minimizing the energy of the system with respect to the
profile function (self-consistent field)
\begin{align}
\frac{\delta E[U_{\rm cl}]}{\delta P(r)} = 0.
\label{energy_minimum}
\end{align}
The profile function at the minimum has a radial size of $\sim 1 \, M^{-1}$.  In leading order of
$1/N_c$ the baryon mass is given by the energy at the minimum (``soliton mass''),
\begin{align}
M_{\rm sol} = E[U_{\rm cl}](\textrm{min}).
\label{m_soliton}
\end{align}

The diagonalization of the Hamiltonian is performed by confining the system in a finite spherical
box and expanding the single-particle wave functions in the eigenfunctions of the vacuum Hamiltonian
($U_{\rm cl} = 1$) \cite{Kahana:1984dx,Kahana:1984be}; see Ref.~\cite{Christov:1995vm} for a review.
The minimization of the energy is performed by discretizing the profile function on a radial grid.
Alternatively, it can be done by employing a variational form, e.g.\
\begin{align}
P(r)= 2\arctan{\frac{R^2}{r^2}},
\label{arctan}
\end{align}
where the parameter $R \sim M^{-1}$ represents the size of the soliton \cite{Diakonov:1987ty} (this
form applies to the chiral limit $M_\pi = 0$; for the extension to $M_\pi \neq 0$ see e.g.\
Ref.~\cite{Schweitzer:2012hh}). With the variational profile one can also explore the behavior of
the system in configurations with ``small'' and ``large'' soliton size. While not corresponding to
solutions of the equations of motion, these configurations allow one to establish a connection with
the quark model and skyrmion pictures of the nucleon and recover both as limiting cases of the
mean-field picture described here \cite{Praszalowicz:1995vi}. This technique is applied to the study
of spin-orbit correlations in Sec.~\ref{subsec:quarkmodel_skyrmion}.

The spatial size of the classical chiral field is of order $M^{-1}$, and the energy of the quark
modes participating in the self-consistent binding is of order $M$.  The sum over energy levels in
Eq.~(\ref{energy_sum}) depends logarithmically on the upper limit.  In the effective dynamics
derived from instantons, the cutoff is provided by the finite size $\bar\rho$ of the zero modes. In
applications to the nucleon and its observables, the cutoff is implemented through a UV
regularization \cite{Christov:1995vm}. This is justified by the fact that logarithmically divergent
sums over the quark levels are dominated by energies $E_n \sim M$ and are generally not sensitive to
the details of UV regularization. In the present study we use the Pauli-Villars regularization,
which preserves the analytic properties of the correlation functions
\cite{Diakonov:1996sr,Diakonov:1997vc}.\footnote{The Pauli-Villars regularization is used in
particular in studies of light-cone correlation functions and partonic structure in the effective
theory, where analytic properties play an essential role (completeness, positivity); see
Refs.~\cite{Diakonov:1996sr,Diakonov:1997vc} for a discussion.}
\subsection{Zero mode quantization}
The self-consistent mean-field solution has no definite momentum or spin/isospin.
Instead, it has zero modes resulting from the invariance of the action under translations 
and rotations/isorotations of the mean field. 
Baryon states with definite momentum and spin/isospin are obtained by performing the 
functional integral over the zero modes.
This is accomplished by parametrizing the zero modes by collective coordinates and 
computing the functional integral over trajectories in these coordinates.
The problem is equivalent to quantizing the translational motion of a point particle
and the rotational motion of a rigid rotor in quantum mechanics and can be solved
using canonical methods \cite{Diakonov:1987ty,Zahed:1986qz}.

The integration over the zero modes is performed in the $1/N_c$ expansion.  The inertial parameters
of the motion are $\mathcal{O}(N_c)$ (mass, moment of inertia), and the velocities of the collective
motion are $\mathcal{O}(N_c^{-1})$ (translational velocity, angular velocity). In leading order of
$1/N_c$ the time dependence of collective coordinates can be neglected, and the functional integral
becomes an ordinary integral over the group manifold of the collective coordinates. In subleading
order of $1/N_c$, the time dependence of the collective coordinates must be retained, and the
collective velocities $\mathcal{O}(N_c^{-1})$ gives rise to dynamical effects.

The translational zero modes are parametrized by shifting the center of the mean field
\begin{subequations}
\label{translations}
\begin{align}
U_{\rm cl}(\bm{x}) &\rightarrow U_{\rm cl}(\bm{x} - \bm{X}),
\\
\Phi_n (\bm{x}) &\rightarrow  \Phi_n(\bm{x} - \bm{X}),
\end{align}
\end{subequations}
where $\bm{X}$ is the center coordinate. Baryon states with momentum $\bm{p} = \mathcal{O}(N_c^0)$
are described by plane-wave wave functions in the coordinate $\bm{X}$,
\begin{align}
e^{i \bm{p}\cdot\bm{X}} .
\end{align}
In leading order of $1/N_c$ the matrix element of a local operator $O(\bm{x})$ (at time $t = 0$)
between baryon states with momenta $\bm{p}$ and $\bm{p}'$ is computed as
\begin{align}
\langle \bm{p}'| O(\bm{0}) | \bm{p} \rangle
&= \int d^3 X \; e^{i(\bm{p}' - \bm{p}) \cdot \bm{X}} \; \langle O(\bm{0}) \rangle_{\bm{X}},
\label{projection_translation}
\end{align}
where $\langle ... \rangle_{\bm{X}}$ denotes the expectation value of the operator in the mean-field
solution centered at $\bm{X}$. Because of translational invariance, this is the same as the
expectation value of the operator at position $\bm{x} = -\bm{X}$ in the mean-field solution centered
at $\bm{X} = 0$. Changing the integration variable to $\bm{x} = -\bm{X}$ one obtains the equivalent
representation,
\begin{align}
\langle \bm{p}'| O(\bm{0}) | \bm{p} \rangle
&= \int d^3 x \; e^{-i(\bm{p}' - \bm{p}) \cdot \bm{x}} \; \langle O(\bm{x}) \rangle_{\bm{X} = 0} ,
\label{projection_translation_alt}
\end{align}
where the translation is now performed in the operator and the mean field remains centered at the
origin.

The rotational zero modes are parametrized by performing a flavor rotation
\begin{subequations}
\label{rotations}
\begin{align}
U_{\rm cl}(\bm{x}) &\rightarrow R \, U_{\rm cl}(\bm{x}) \, R^\dagger ,
\\
\Phi_n (\bm{x}) &\rightarrow  R \, \Phi_n(\bm{x}),
\end{align}
\end{subequations}
where $R$ is an SU(2) rotation matrix (we consider the case of two flavors, see Sec.~\ref{sec:qcd}).
Because of the hedgehog symmetry of the mean field, Eq.~(\ref{hedgehog}), such a flavor rotation is
equivalent to a spatial rotation. Baryon states with spin $\{S, S_3\}$ and isospin $\{ T, T_3 \}$
are described by rotational wave functions in the variable $R$. The hedgehog symmetry of the mean
field implies that the physical states occur in representations obeying the constraint
$\bm{S}+\bm{T}=0$ (this constraint realizes the spin-flavor symmetry of baryons in the large-$N_c$
limit in the mean-field picture).  The rotational wave functions in these representations are given
by
\begin{align}
\phi^{S=T}_{S_3T_3}(R) &=  \sqrt{2S+1} \, (-1)^{S+S_3} \, D^{S=T}_{T_{3}, -S_{3}} (R),
\end{align}
where $D$ is the SU(2) Wigner $D$ function \cite{Landau:1991wop}. In leading order of $1/N_c$, the
matrix element of an operator between baryon states with spin-isospin $S = T$ and projections
$\{ S_3, T_3 \}$ and $\{ S_3', T_3' \}$ is computed as
\begin{align}
& \langle S=T, S_3', T_3' | \, O (\bm{0}) \, | S=T, S_3, T_3\rangle
\nonumber \\
&= \int dR\;\phi^{\ast\;S=T}_{S_3' T_3'}(R) \; \phi^{S=T}_{S_3T_3}(R) \;
\langle O(\bm{0}) \rangle_{R},
\label{projection_rotation}
\end{align}
where the integration is over the SU(2) group volume $(\int dR = 1)$ and
$\langle O(\bm{0}) \rangle_{\bm{R}}$ denotes the average in the rotated mean field.

In subleading order of $1/N_c$, effects of the finite rotational velocity must be included.  The
functional integral over time-dependent rotations $R(t)$ is computed using the canonical
quantization formalism \cite{Christov:1995vm}, applying the quantization rule
\begin{align}
R^\dagger \frac{dR}{dt} &\rightarrow \frac{i}{2I} S^a\tau^a,
\label{quantization-rule}
\end{align}
where $S^a$ are the spin operators and $\tau^a$ the isospin Pauli matrices. The constant
$I= \mathcal{O}(N_c)$ is the moment of inertia of the soliton, which is obtained as a sum over quark
single-particle levels similar to the soliton mass [see Eq.~(\ref{energy_sum})]; see
Ref.~\cite{Christov:1995vm} for its explicit expression.  At this order the energy of the baryon
states is then described by the collective Hamiltonian
\begin{align}
H_{\mathrm{coll}} &= M_{\mathrm{sol}} + \frac{\bm{S}^2}{2I}.
\end{align}
The spin-dependent term is $\mathcal{O}(N_c^{-1})$ and gives rise to a mass splitting between the
$N$ and $\Delta$ states ($S = 1/2$ and $3/2$).

In the following analysis the translational motion is treated in leading order in $1/N_c$;
corrections resulting from the translational velocity can be neglected.  The rotational motion is
treated in leading or next-to-leading order in $1/N_c$ (lowest nonvanishing order), as required by
the quantum numbers of the operators being evaluated.
\subsection{Matrix elements of effective operators}
\label{subsec:matrix_elements}
%
%
\begin{figure}[t]
\centering
\includegraphics[width=0.8\columnwidth]{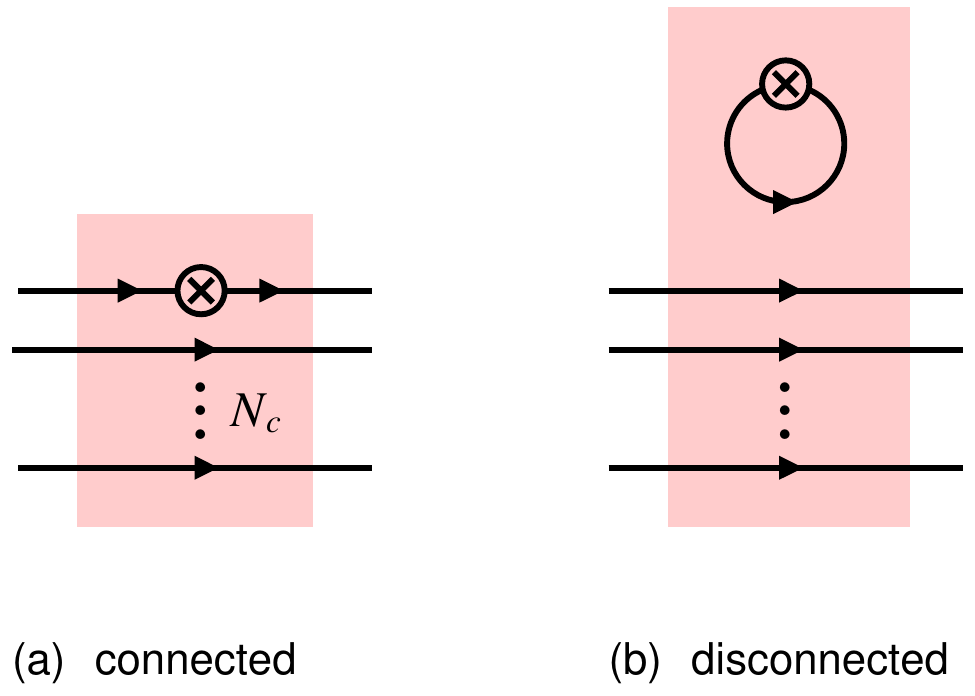}
\caption{Matrix element of a quark one-body operator in the nucleon in the effective theory (see
Fig.~\ref{fig:baryon}). (a)~Connected diagrams. (b)~Disconnected diagrams. The terms connected and
disconnected refer to quark lines in the background of the classical chiral field.}
\label{fig:baryon_operator}
\end{figure}
Matrix elements of effective operators between nucleon (and other baryon) states in the effective
theory are computed in the $1/N_c$ expansion
\cite{Diakonov:1987ty,Wakamatsu:1990ud,Christov:1995vm,Diakonov:1996sr}.  In the present study we
are dealing with local quark bilinear operators of the form
\begin{align}
O(x) = \bar\psi (x) \, \Gamma [U, \partial] \, \psi (x),
\label{operator_general}
\end{align}
where $\Gamma$ is a kernel that may involve derivatives of the quark fields and depend on the
classical chiral field; $\Gamma$ is a spinor-flavor matrix, and the summation over color, spinor,
and flavor indices is implied (see Sec.~\ref{subsec:effective_operators}).  The matrix elements are
extracted from the Euclidean three-point correlation function, evaluated in the saddle point
approximation. In the first step one computes the average over the fermion fields in the background
of the classical chiral field.  The fermionic integral includes contractions where the operator is
connected to the baryon through quark lines attached to the source/sink (``connected diagrams''),
and where the operator appears in a closed quark loop but is connected with the baryon through the
chiral field (``disconnected diagrams''), see Figure~\ref{fig:baryon_operator}.  The contractions
are evaluated by expanding the quark propagators in the single-particle wave functions
Eq.~(\ref{spectrum}), converting the second-quantized to a first-quantized average. The fermion
average is obtained as [for the classical chiral field centered at the origin, see
Eq.~(\ref{hedgehog})]
\begin{align}
\langle O(\bm{0})\rangle = N_c \langle \textrm{lev} | \Gamma | \textrm{lev} \rangle
+ N_c \sum_{n \; \textrm{neg.cont}} \langle n | \Gamma | n \rangle .
\label{matrix_element_sum}
\end{align}
The first term is the contribution of the discrete level; the second term is the contribution of the
negative continuum levels (see Fig.~\ref{fig:baryon}c). In both cases the levels are occupied by
$N_c$ quarks, and the average gives rise to a factor $N_c$. The presence of the continuum
contribution is an essential feature of the field-theoretical description of the nucleon in the
large-$N_c$ limit. Both contributions together are needed to satisfy conservation laws, sum rules,
and other features relying on the completeness of states \cite{Diakonov:1996sr}.

In the second step one performs the integral over the zero modes of the classical chiral field and
projects on the baryon states with definite momentum and spin-isospin quantum numbers, denoted
collectively as
\begin{align}
B \equiv \{S = T, S_{3}, T_{3}\},
\hspace{1em}
B' \equiv \{S' = T', S_{3}', T_{3}' \}.
\label{spinflavor_B}
\end{align}
In leading order of $1/N_c$ the angular velocity of the collective rotations can be neglected; the
integration is over static rotations; and the projection is performed with
Eq.~(\ref{projection_rotation}).  Baryon matrix elements appearing at this order are of the generic
form
\begin{align}
\langle B'| O(\bm{0}) |B\rangle = \langle X \rangle_{B'B} \; \times \;
N_c \sum_{n \; \textrm{occ}} \langle n | \Gamma | n \rangle .
\label{single_sum}
\end{align}
Here $X$ denotes an operator in the spin-flavor quantum number, whose specific form comes out of the
quantization procedure and depends on the spin-flavor quantum numbers of $O$.  The label ``occ''
denotes the summation over occupied quark single-particle levels, including the discrete level and
the negative continuum [see Eq.~(\ref{matrix_element_sum}) and Fig.~\ref{fig:baryon}c].  Such matrix
elements are given by single sums over quark levels.  In next-to-leading order of $1/N_c$ the
angular velocity of the collective rotations must be retained; the integration is a functional
integral over time-dependent rotations. It is performed by expanding $\langle O(\bm{0})\rangle_R$ to
first order in the velocity $dR/dt$ and applying the canonical quantization prescription
Eq.~(\ref{quantization-rule}).  Baryon matrix elements appearing at this order are of the generic
form
\begin{align}
\langle B'| O(\bm{0}) |B\rangle &= \langle X^a \rangle_{B'B}
\nonumber \\
& \times \frac{N_c}{I} \!\!
\sum_{ \substack{n \; \mathrm{occ} \\ m \; \mathrm{non-occ} }} \!\!
\frac{\langle n | \tau^{a} | m \rangle \langle m | \Gamma | n \rangle}{E_n - E_m} ,
\label{double_sum}
\end{align}
where $X^a$ is a spin-flavor operator. Such matrix elements are given by double sums over quark
levels, describing particle-hole excitations of the fermionic ground state in the mean field (see
Fig.~\ref{fig:baryon}c). The suppression in $1/N_c$ arises from the inverse of the moment inertia
$I = \mathcal{O}(N_c)$ in Eq.~(\ref{double_sum}).
\section{Spin-orbit correlations in nucleon}
\label{sec:spinorbit}
\subsection{$N_{c}$ scaling}
The mean-field picture of baryons and the effective twist-3 operators from instantons can be
combined to compute the nucleon matrix elements of the QCD operator Eq.~(\ref{operator_def}) and
study the properties of the quark spin-orbit correlations. We first derive the $N_c$ scaling of the
spin-flavor components of the nucleon matrix element and the form factors of
Eq.~(\ref{parametrization}). The scaling behavior is model-independent and represents a consequence
of the spin-flavor symmetry of baryons in the large-$N_c$ limit.  We derive it here from the
specific realization in the mean-field picture, which will be used in the subsequent dynamical
calculations. The extension to $N$-$\Delta$ transition matrix elements of the operator
Eq.~(\ref{operator_def}) is discussed in Sec.~\ref{subsec:ndelta}.

In the large-$N_c$ limit the baryon masses are $M_{N, \Delta} = \mathcal{O}(N_{c})$, and the
$N$-$\Delta$ mass splitting is $M_\Delta - M_N = \mathcal{O}(N_c^{-1})$.  The $1/N_c$ expansion of
transition matrix elements of operators is performed in a class of reference frames where the
initial and final baryon 3-momenta $|\bm{p}|$ and $|\bm{p}'|$ are $\mathcal{O}(N_c^0)$.  In such
frames the components of the average 4-momentum and the 4-momentum transfer,
Eq.~(\ref{average_difference_momenta}), scale as
\begin{align}
&\Delta^0 = \mathcal{O}(N^{-1}_c), \hspace{1em} \bm{\Delta} = \mathcal{O}(N^0_c), 
\\[1ex]
&P^0 = \mathcal{O}(N_c), \hspace{1em} \bm{P} = \mathcal{O}(N^0_c) .
\end{align}
$P^0$ is equal to the static baryon mass (the soliton mass $M_{\rm sol}$ in the mean-field approach)
up to corrections $\mathcal{O}(N_c^{-1})$ and can be set to the nucleon mass,
$P^0 = M_N + \mathcal{O}(N_c^{-1})$.  The invariant momentum transfer is
\begin{align}
t = -|\bm{\Delta}|^2 + \mathcal{O}(N_c^{-1}).
\end{align}
The spin-flavor quantum numbers are denoted as in Eq.~(\ref{spinflavor_B}), and the baryon states
are normalized as
\begin{align}
\langle B', \bm{p}' | B, \bm{p} \rangle
= 2M_N (2\pi)^3 \delta^{(3)}(\bm{p}'-\bm{p}) \delta_{B'B} ,
\end{align}
where taking the baryon mass as the nucleon mass is correct in leading and next-to-leading order in
$1/N_c$.

It is natural to work in the symmetric frame $\bm{P} = 0$ (Breit frame), where $\bm{\Delta}$ is the
only 3-vector and the transitions obey simple spin selection rules. The transition matrix element of
the tensor operator Eq.~(\ref{operator_def}) in this frame is denoted by
\begin{align}
\langle B'| T^{\mu \nu}_{5}| B\rangle
\equiv \langle B', \bm{\Delta}/2| T^{\mu \nu}_{5}(0) | B, -\bm{\Delta}/2 \rangle .
\end{align}
The $1/N_c$ expansion is performed by computing the transition matrix elements in the mean-field
approach.  The zero mode quantization procedure of Sec.~\ref{subsec:matrix_elements} is applied in
abstract form, using only general properties of the mean-field expectation values of the operator,
but not assuming any definite values (such values will be obtained later from the specific dynamics
of the effective theory).  In this way we obtain the following results for the transition matrix
elements of the 3-dimensional components of the tensor operator in the isoscalar and isovector
channels [see Eq.~(\ref{operator_isospin})]:
\begin{subequations}
\label{matrixelement_largenc}
\begin{align}
\langle B' |(T^{S}_{5})^{0k} |B\rangle
&= -4 M^{2}_{N} \langle  S^{k}  \rangle_{B'B} M^{S}_{0}(t) + ...,
\\[1ex]
\langle B' |(T^{V}_{5})^{0k} |B\rangle 
&= -4 M^{2}_{N}  \langle  D^{3k}  \rangle_{B'B} M^{V}_{0}(t) + ..., 
\\[1ex]
\langle B' |(T^{S}_{5})^{ \{0k\} } |B\rangle 
&= -4 M^{2}_{N} \langle  S^{k}  \rangle_{B'B} \bar{M}^{S}_{0}(t) + ..., 
\\[1ex]
\langle B' |(T^{V}_{5})^{ \{0k\} } |B\rangle 
&= -4 M^{2}_{N} \langle  D^{3k}  \rangle_{B'B} \bar{M}^{V}_{0}(t) + ...,  \hspace{-1em}
\\[1ex]
\langle B' |(T^{S}_{5})^{[ij]} |B\rangle 
&=  -2 M_{N} i \epsilon^{jik} \Delta^{k}
\nonumber \\
& \times 1_{B'B} \, M^{S}_{1}(t) + ...,
\\[1ex]
\langle B' |(T^{V}_{5})^{[ij]} |B\rangle 
&=  -2 M_{N} i \epsilon^{jik} \Delta^{k} 
\nonumber \\
& \times
\langle \{ S^a, D^{3a}\} \rangle_{B'B} \, M^{V}_{1}(t) + ...,
\end{align}
\end{subequations}
where $...$ denotes higher multipole structures which are irrelevant to the present study.  The
spin-flavor dependence is determined by the matrix elements of the spin-flavor operators coming out
of the collective quantization procedure, Eq.~(\ref{projection_rotation}), which are given by SU(2)
Clebsch-Gordan coefficients \cite{Christov:1995vm}.  The dynamical information in
Eq.~(\ref{matrixelement_largenc}) is contained in the multipole form factors
$M^{S}_{0}(t)$--$M^{V}_{1}(t)$. The mean-field picture predicts their $N_c$ scaling as
\begin{subequations}
\label{multipole_ncscaling}
\begin{align}
M^{S}_{0}(t) &= \mathcal{O}(N^{-1}_{c}),
\hspace{2em}
M^{V}_{0}(t) = \mathcal{O}(N^{0}_{c}),
\\[1ex]
\bar{M}^{S}_{0}(t) &= \mathcal{O}(N^{-1}_{c}), 
\hspace{2em}
\bar{M}^{V}_{0}(t) = \mathcal{O}(N^{0}_{c}), 
\\[1ex]
M^{S}_{1}(t) &= \mathcal{O}(N^{1}_{c}),
\hspace{2em}
M^{V}_{1}(t) = \mathcal{O}(N^{0}_{c}).
\end{align}
\end{subequations}
In the calculation in the effective theory, $M^{V}_{0}$, $\bar{M}^{V}_{0}$, and $M^{S}_{1}$ are
obtained as single sums over quark levels, Eq.~(\ref{single_sum}), while
$M^{S}_{0}, \bar{M}^{S}_{0}$ and $M_1^V$ are obtained as double sums, Eq.~(\ref{double_sum}); the
explicit expressions are not needed here and will be presented elsewhere~\cite{kk}.
Equations~(\ref{matrixelement_largenc}) and (\ref{multipole_ncscaling}) predict the spin-flavor
structure and $N_c$ scaling of the baryon matrix elements for all transitions between $S=T$ states,
including $N$ and $\Delta$ states.

To obtain the $N_c$ scaling of the nucleon matrix elements, we evaluate the large-$N_c$ expressions
of Eq.~(\ref{matrixelement_largenc}) for transitions between nucleon states and match them with the
general multipole expansion of the nucleon transition matrix elements of the tensor operator.  We
denote the spin components of the nucleon matrix elements in the symmetric frame by
\begin{align}
\langle S_3'| T^{\mu \nu}_{5} |S_{3} \rangle \equiv
\langle N (\bm{\Delta}/2, S_3') |T_{5}^{\mu\nu}| N (-\bm{\Delta}/2, S_3) \rangle
\end{align}
(we only exhibit the spin structure here; the isospin structure follows the standard selection
rules).  Performing the multipole expansion of the matrix elements of the 3-dimensional tensor
components in the symmetric frame, and expressing them in terms of the invariant form factors of
Eq.~(\ref{parametrization}), we obtain
\begin{subequations}
\label{matrixelement_nucleon}
\begin{align}
\langle S_3'| T^{0k}_{5} |S_{3} \rangle
&= - 4 M^{2}_{N} S_{S_{3}^{\prime}S_{3}}^{k} 
\nonumber \\
& \times \left[ \frac{1}{2} \tilde{A}(t) + \frac{1}{2} \tilde{C}(t) - \tilde{F}(t) \right] + ...,
\\[1ex]
\langle S_3'| T^{ \{0k\}}_{5} |S_{3} \rangle
&= - 4 M^{2}_{N} S_{S_{3}^{\prime}S_{3}}^{k} \tilde{A}(t) + ...,
\\[1ex]
\langle S_3'| T^{[ij]}_{5} |S_{3} \rangle
&= -  2 M_{N} i \epsilon^{jik} \Delta^{k} \delta_{S_{3}^{\prime}S_{3}} \tilde{F} (t),
\end{align}
\end{subequations}
where $...$ denotes higher multipole structures that are not needed in the following.
$S_{S_{3}^{\prime}S_{3}}^k$ is the matrix element of the spin operator between $S = 1/2$ states; its
spherical vector components ($M = -1, 0, 1$) are given by
\begin{align}
S^{M}_{S'_{3}S_{3}} = \frac{\sqrt{3}}{2} \, C^{\frac{1}{2} S'_{3} }_{\frac{1}{2} S_{3} 1M},
\end{align}
and the Cartesian components ($k = x, y, z$) are obtained from the spherical ones as
\begin{subequations}
\begin{align}
&S^{x} = \frac{1}{\sqrt{2}} (S^{-1}- S^{+1}), \\
&S^{y} = \frac{i}{\sqrt{2}} (S^{-1}+ S^{+1}), \\
&S^{z} =  S^{0}.
\end{align}
\end{subequations}
The multipoles in Eq.~(\ref{matrixelement_nucleon}) are now matched with the large-$N_c$ expressions
Eq.~(\ref{matrixelement_largenc}). For transitions between nucleon states
$(S = T = 1/2, S' = T' = 1/2)$ the matrix elements of the spin-flavor operators in
Eq.~(\ref{matrixelement_largenc}) are given by
\begin{subequations}
\label{S_D_matrix_elements}
\begin{align}
&\langle S^{i} \rangle_{B'B} \; = \; S^{i}_{S_3' S_3} \, \delta_{T_3' T_3},
\\[1ex]
&\langle D^{3i} \rangle_{B'B} \; = \; -\frac{2}{\sqrt{3}} \,
C^{\frac{1}{2} T_{3}'}_{\frac{1}{2} T_{3} 1 0} \, S^{i}_{S_3' S_3} ,
\\[1ex]
&\langle \{S^{i}, D^{3j} \} \rangle_{B'B} \; = \; -\frac{1}{\sqrt{3}} \, \delta_{S_3' S_3} \,
C^{\frac{1}{2} T_{3}'}_{\frac{1}{2} T_{3} 1 0} \, \delta^{ij}.
\end{align}
\end{subequations}
In this way we can express the invariant form factors in terms of the large-$N_c$ multipole form
factors and predict their $N_c$ scaling. In the isoscalar channel we obtain\footnote{Here we limit
ourselves to performing the matching at $t = 0$. For $t \neq 0$ the higher multipoles in
Eq.~(\ref{matrixelement_nucleon}) need to be taken into account and give terms proportional to $t$
in the form factor relations. This would be necessary in order to obtain the relations for the form
factors $\tilde{B}$ and $\tilde{D}$ of Eq.~(\ref{parametrization}).}
\begin{subequations}
\label{invariant_from_multipole_isoscalar}
\begin{align}
&\tilde{A}^{u+d}(0) =  \bar{M}^{S}_{0}(0) ,
\\[1ex]
&\tilde{F}^{u+d}(0)= M^{S}_{1}(0)  ,
\\[1ex]
&\tilde{C}^{u+d}(0) =  2 M^{S}_{0}(0) - \bar{M}^{S}_{0}(0)  + 2M^{S}_{1}(0).
\end{align}
\end{subequations}
In the isovector channel we obtain
\begin{subequations}
\label{invariant_from_multipole_isovector}
\begin{align}
&\tilde{A}^{u-d}(0) = -\frac{2}{3}  \bar{M}^{V}_{0}(0), 
\\[1ex]
&\tilde{F}^{u-d}(0)= M^{V}_{1}(0),
\\[1ex]
&\tilde{C}^{u-d}(0) = - \frac{4}{3} M^{V}_{0}(0) +\frac{2}{3} \bar{M}^{V}_{0}(0)  + 2M^{V}_{1}(0). 
\end{align}
\end{subequations}
The $N_{c}$ scaling of the invariant form factors is then obtained from
Eq.~(\ref{multipole_ncscaling}) as
\begin{subequations}
\begin{align}
& \tilde{A}^{u+d}(0) = \mathcal{O}(N^{-1}_{c}),
\hspace{1.5em}
\tilde{A}^{u-d}(0) = \mathcal{O}(N^{0}_{c}), 
\\
& \tilde{C}^{u+d}(0) = \mathcal{O}(N^{1}_{c}),
\hspace{2em}
\tilde{C}^{u-d}(0) = \mathcal{O}(N^{0}_{c}), 
\\
& \tilde{F}^{u+d}(0)= \mathcal{O}(N^{1}_{c}),
\hspace{2em}
\tilde{F}^{u-d}(0)= \mathcal{O}(N^{0}_{c});
\end{align}
\end{subequations}
this scaling behavior extends to the form factors at $t \neq 0$. Furthermore,
Eqs.~(\ref{multipole_ncscaling}) and (\ref{invariant_from_multipole_isoscalar}) imply that in
leading order of the $1/N_c$ expansion
\begin{align}
\tilde{F}^{u+d}(t) = \frac{1}{2} \tilde{C}^{u+d}(t);
\end{align}
this relation is valid at $t \neq 0$.  The isoscalar spin-orbit correlations in
Eq.~(\ref{spinorbit_def}) can therefore equivalently be expressed in terms of $\tilde{F}^{u+d}(0)$
instead of $\tilde{C}^{u+d}(0)$.  Altogether, in leading order of the $1/N_c$ expansion we obtain
for the spin-orbit correlations
\begin{subequations}
\label{spinorbit_formfactor_largenc}
\begin{align}
C^{u+d}_{z} &= \frac{1}{2} \tilde{C}^{u+d}(0) + \mathcal{O}(N^{-1}_{c}) 
\label{spinorbit_largenc_isoscalar}
\nonumber \\[1ex]
&=  \tilde{F}^{u+d}(0) + \mathcal{O}(N^{-1}_{c}),
\\[1ex]
C^{u-d}_{z} &= \frac{1}{2} \left[ \tilde{A}^{u-d}(0) + \tilde{C}^{u-d}(0) \right].
\end{align}
\end{subequations}
One observes that in the isoscalar channel the dominant contribution comes from the antisymmetric
part of the tensor operator (form factors $\tilde{C}^{u+d}$ or $\tilde{F}^{u+d}$); the contribution
of the symmetric part is suppressed (form factor $\tilde{A}^{u+d}$).  In the isovector channel the
symmetric and antisymmetric parts of the tensor contribute at the same order. This can also be seen
using the QCD equation-of-motion relation Eq.~(\ref{formfactors_from_vector}) and the known $N_c$
scaling of the vector form factors,
\begin{subequations}
\begin{align}
& \frac{1}{2} \left[\tilde{C}^{u+d}(0) + \tilde{A}^{u+d}(0)\right] 
\nonumber \\[1ex]
& = \underbrace{  -\frac{1}{2}F^{u+d}_{1}(0)}_{\mathcal{O}(N^{1}_{c})}
  + \underbrace{ \frac{1}{2} \tilde{A}^{u+d}(0)}_{\mathcal{O}(N^{-1}_{c})}, 
\\[1ex]
& \tilde{F}^{u+d}(0) = \underbrace{ -\frac{1}{2}F^{u+d}_{1}(0)}_{\mathcal{O}(N^{1}_{c})} .
\end{align}
\end{subequations}
Equation~(\ref{spinorbit_formfactor_largenc}) thus implies the following $N_{c}$ scaling for the
spin-orbit correlation 
\begin{align}
C^{u+d}_{z} = \mathcal{O}(N^{1}_{c}), \hspace{2em} C^{u-d}_{z} = \mathcal{O}(N^{0}_{c}).
\end{align}
The isoscalar spin-orbit correlation is parametrically larger than the isovector one, which agrees
with the findings of Refs.~\cite{Lorce:2014mxa,Lorce:2011kd}. Note that the subleading contributions
to the isoscalar spin-orbit correlation are suppressed by $1/N^{2}_{c}$ relative to the leading
contribution.\footnote{This statement applies only to the $1/N_c$ corrections coming out of the
rotational zero mode quantization considered here. It does not apply to $1/N_c$ corrections arising
from quantum fluctuations of the chiral fields.}
\subsection{First-quantized representation}
\label{subsec:first_quantized}
We now compute the isoscalar spin-orbit correlation in the nucleon,
Eq.~(\ref{spinorbit_largenc_isoscalar}), which appears in leading order of the $1/N_{c}$ expansion.
In the mean-field picture the nucleon matrix element is obtained as a sum of matrix elements in
quark single-particle levels in the classical chiral field, see Sec.~\ref{subsec:matrix_elements}.
This allows us to see how the spin-orbit correlation operator is expressed in first-quantized form,
and how the effective equation of motion is realized at the level of quark single-particle levels.
It also allows us to discuss the role of the kinetic and potential terms in the effective operator.

The isoscalar spin-orbit correlation in the large-$N_c$ limit can be obtained by computing either
the form factor $\tilde C^{u + d}$ or the form factor $\tilde F^{u + d}$; the two form factors
coincide in the large-$N_c$ limit; see Eq.~(\ref{spinorbit_largenc_isoscalar}). We compute the form
factor $\tilde F^{u + d}(t)$, using the effective operator of Eq.~(\ref{effective_operator_twist3}),
the form of the matrix element of Eq.~(\ref{matrixelement_nucleon}c), and the prescription for the
calculation of nucleon matrix elements in Sec.~\ref{subsec:matrix_elements}.  From the kinetic term
of the effective operator Eq.~(\ref{antisymmetric_kin}) we obtain
\begin{align}
\tilde{F}^{u+d}(t)[\textrm{kin}]
&= i N_c \, \epsilon^{ijk} \frac{\Delta^k}{|\bm{\Delta}|^2} \int d^3 x \;
e^{i\bm{\Delta}\cdot\bm{x}}
\nonumber \\
& \times
\sum_{n \; \mathrm{occ}} \Phi^{\dagger}_{n}(\bm{x}) \;
i \gamma^0 \gamma^i \gamma_5 \overleftrightarrow{\partial}_{\!\! j} \; \Phi_{n}(\bm{x}),
\label{Ftilde_kin_Delta}
\end{align}
where ``occ'' means the sum over occupied single-particle levels, including the discrete level and
the negative continuum; see Eq.~(\ref{single_sum}). Because the expression depends only on
$t = -|\bm{\Delta}|^2$, we can average the right-hand side over the direction of the vector
$\bm{\Delta}$, using
\begin{align}
\int \frac{d\Omega_\Delta}{4\pi} \; \frac{\Delta^k}{|\bm{\Delta}|^2} e^{i\bm{\Delta}\cdot\bm{x}}
= i x^k \; \frac{j_{1}(|\bm{x}| |\bm{\Delta}|)}{|\bm{x}| |\bm{\Delta}|},
\label{average_direction_Delta}
\end{align}
where $j_1$ is the spherical Bessel function. The coordinate vector appearing here can be combined
with the derivative vector in Eq.~(\ref{Ftilde_kin_Delta}) to form the first-quantized orbital
angular momentum operator
\begin{align}
\bm{L} \equiv \bm{x} \times \bm{p}, \hspace{1em} \bm{p}
\equiv \frac{i}{2}(\overleftarrow{\bm{\nabla}} - \overrightarrow{\bm{\nabla}}) ,
\end{align}
where $\bm{\nabla}$ denotes the gradient vector.  The Dirac matrices in Eq.~(\ref{Ftilde_kin_Delta})
can be combined to give the spin operator for Dirac spinors \cite{Berestetskii:1982qgu},
\begin{align}
\frac{1}{2} \Sigma^i \equiv -\frac{1}{2} \gamma^0 \gamma^i \gamma_5 .
\end{align}
Altogether the form factor Eq.~(\ref{Ftilde_kin_Delta}) becomes
\begin{align}
&\tilde{F}^{u+d}(t)[\textrm{kin}]
\nonumber \\
&= N_c \int d^3 x \; \frac{j_{1}(|\bm{x}| |\bm{\Delta}|)}{|\bm{x}| |\bm{\Delta}|}
\sum_{n \; \mathrm{occ}}
\Phi^{\dagger}_{n}(\bm{x}) \; \bm{L} \cdot \bm{\Sigma} \; \Phi_{n}(\bm{x}) .
\label{Ftilde_kin_sum}
\end{align}
The spin-orbit correlation is given by the value at zero momentum transfer,
\begin{align}
C^{u+d}_{z}[\textrm{kin}] &= \tilde{F}^{u+d}(0)[\textrm{kin}]
\nonumber \\
&= \frac{N_c}{3} \int d^3 x \; 
\sum_{n \; \mathrm{occ}} \;
\Phi^{\dagger}_{n}(\bm{x}) \; \bm{L} \cdot \bm{\Sigma} \; \Phi_{n}(\bm{x}) .
\label{spinorbit_kin_sum}
\end{align}
One sees that the nucleon matrix element of the kinetic term of the effective operator describes the
correlation between the quark spin and orbital angular momentum in the first-quantized
representation.

From the potential term of the effective operator Eq.~(\ref{antisymmetric_pot}) we obtain
\begin{align}
&\tilde{F}^{u+d}(t)[\textrm{pot}]
= \frac{i N_c}{2} \, \epsilon^{ijk} \frac{\Delta^k}{|\bm{\Delta}|^2}
\int d^3 x \; e^{i\bm{\Delta}\cdot\bm{x}}
\nonumber \\
& \times
\sum_{n \; \mathrm{occ}} \Phi^{\dagger}_{n}(\bm{x}) \;
i \gamma^0 \sigma^{ij} \gamma_5 M U^{\gamma_5} (\bm{x}) \; \Phi_{n}(\bm{x}) .
\label{Ftilde_pot_Delta}
\end{align}
We again average over the directions of $\bm{\Delta}$ as in Eq.~(\ref{average_direction_Delta}).
The spin operator now arises from the contraction
\begin{align}
\frac{1}{2} \Sigma^k = \frac{1}{4} \epsilon^{ijk} \sigma^{ij} .
\end{align}
The form factor Eq.~(\ref{Ftilde_pot_Delta}) becomes
\begin{align}
&\tilde{F}^{u+d}(t)[\textrm{pot}]
= - N_c \int d^3 x \; \frac{j_{1}(|\bm{x}| |\bm{\Delta}|)}{|\bm{x}| |\bm{\Delta}|}
\nonumber \\
&\times \sum_{n \; \mathrm{occ}} \Phi^{\dagger}_{n}(\bm{x}) \;
i \, \bm{x} \cdot \bm{\Sigma} \; \gamma^0 \gamma_5 M U^{\gamma_5} (\bm{x})
\Phi_{n}(\bm{x}) .
\end{align}
The spin-orbit correlation is again given by the value at zero momentum transfer,
\begin{align}
&C^{u+d}_{z}[\textrm{pot}] = \tilde{F}^{u+d}(0)[\textrm{pot}]
\nonumber \\
&= - \frac{N_c}{3} \int d^3 x \;
\sum_{n \; \mathrm{occ}} \Phi^{\dagger}_{n}(\bm{x}) \;
i \, \bm{x} \cdot \bm{\Sigma} \; \gamma^0 \gamma_5 M U^{\gamma_5} (\bm{x})
\Phi_{n}(\bm{x}) ;
\label{spinorbit_pot_sum}
\end{align}
the combination of Dirac matrices appearing here can also be written as
\begin{align}
i \, \bm{x} \cdot \bm{\Sigma} \; \gamma^0 \gamma_5 = - i \, \bm{x} \cdot \bm{\gamma}.
\end{align}
One sees that the nucleon matrix element of the potential term of the effective operator describes
the correlation between the radius vector and the spin of the quarks in the first-quantized
representation. Such a correlation becomes possible because the $\gamma_5$ bilinear form provides a
pseudoscalar density that allows the radius vector to couple to the spin. This remarkable result
represents a relativistic effect and implies a new mechanical interpretation of the spin-orbit
correlations.

The total effective operator Eq.~(\ref{effective_operator_twist3}), including kinetic and potential
terms, obeys the relation Eq.~(\ref{total_derivative_effective}), and the total nucleon matrix
element satisfies the relation
\begin{align}
C^{u+d}_{z} &= C^{u+d}_{z}[\textrm{kin}] + C^{u+d}_{z}[\textrm{pot}]
\nonumber \\
&= - \frac{1}{2} G_E^{u+d}(0) = -\frac{N_c}{2}.
\label{sumrule_effective}
\end{align}
We now want to see how this relation is realized in the first-quantized representation of
Eqs.~(\ref{spinorbit_kin_sum}) and (\ref{spinorbit_pot_sum}).  The Dirac equation for the quark
single-particle wave functions and their complex conjugate are
\begin{subequations}
\label{eom_single}
\begin{align}
\left[ - i \gamma^{0} \gamma^{i} \overrightarrow{\partial}_{\! i}
+ M \gamma^{0} U^{\gamma_{5}}(\bm{x}) \right] \Phi_{n}(\bm{x})
= E_n \Phi_{n}(\bm{x}),
\\[1ex]
\Phi^{\dagger}_{n}(\bm{x}) \left[ i \gamma^{0} \gamma^{i} \overleftarrow{\partial}_{\! i}
+ M \gamma^{0} U^{\gamma_{5}}(\bm{x}) \right]
= \Phi^{\dagger}_{n}(\bm{x}) E_n .
\end{align}
\end{subequations}
Multiplying Eq.~(\ref{eom_single}a) from the left by
$\Phi^\dagger (\bm{x}) \, x^{j} \gamma^{0} \gamma^{j}$, and Eq.~(\ref{eom_single}b) from the right
by $x^{j} \gamma^{0} \gamma^{j} \, \Phi (\bm{x})$, and subtracting the two equations, the terms on
the right-hand side proportional to $E_n$ cancel.  Reducing the product of Dirac matrices,
rearranging the derivatives acting on the wave functions, and summing over occupied single-particle
levels, we obtain
\begin{align}
& \phantom{+} \; \frac{N_{c}}{3} \sum_{n \; \mathrm{occ}}
\Phi^{\dagger}_{n}(\bm{x}) \, \bm{L}\cdot \bm{\Sigma} \, \Phi_{n}(\bm{x}) 
\nonumber \\
& - \frac{N_c}{3}
\sum_{n \; \mathrm{occ}}
\Phi^{\dagger}_{n}(\bm{x}) \; i \, \bm{x} \cdot \bm{\Sigma} \;
\gamma^0 \gamma_5 M U^{\gamma_5} (\bm{x}) \Phi_{n} (\bm{x})
\nonumber \\
= &- \frac{N_{c}}{2}  \sum_{n \; \mathrm{occ}}
\Phi^{\dagger}_{n} (\bm{x}) \Phi_{n} (\bm{x})
\nonumber \\
& + \frac{N_{c}}{6}  \sum_{n \; \mathrm{occ}}
\partial_i \left[ \Phi^{\dagger}_{n} (\bm{x}) x^i \Phi_{n} (\bm{x}) \right] .
\label{sumrule_densities}
\end{align}
The first term on the right-hand side is $-1/2$ times the baryon number density in the large-$N_c$
nucleon.  When integrated over the position, it gives
\begin{align}
&-\frac{N_{c}}{2}  \int d^{3} x  \sum_{n \; \mathrm{occ}}
\Phi^{\dagger}_{n}(\bm{x}) \Phi_{n}(\bm{x}) 
\nonumber \\
&= -\frac{1}{2} G^{u+d}_{E}(0)
= -\frac{N_{c}}{2} .
\label{charge_sum}
\end{align}
The second term on the right-hand side of Eq.~(\ref{sumrule_densities}) is a total derivative and
integrates to zero.  Thus one sees that the sum rule Eq.~(\ref{sumrule_effective}) is satisfied in
the large-$N_c$ nucleon state.  This validates the first-quantized form of the effective operators
and confirms the role of the potential term.

It is worth noting that the sum rule for the isoscalar spin-orbit correlation,
Eq.~(\ref{sumrule_effective}), is satisfied independently in each quark single-particle level. It
requires only the equations of motion of the quarks in the classical background field,
Eq.~(\ref{eom_effective}) or Eq.~(\ref{eom_single}), not the equation of motion of the chiral field
presented by the self-consistency condition Eq.~(\ref{energy_minimum}).  As such it is similar to
the baryon number sum rule relating the integral of the quark minus antiquark densities to the
baryon charge \cite{Diakonov:1996sr,Diakonov:1997vc}.

It is also worth emphasizing that the relation between the densities, Eq.~(\ref{sumrule_densities}),
involves a total derivative term in addition to the baryon number density on the right-hand
side. When integrating over space, the total derivative term drops out (it produces a surface term
at infinity), and the sum rule for the integrated quantities is realized. However, when discussing
spatial densities, the total derivative term is nonzero and must be taken into account. The
formulation of spatial densities of the tensor operator Eq.~(\ref{operator_def}) and their
connection with spin-orbit correlations is left for a future study. The effective operator formalism
and the large-$N_c$ picture of the nucleon can be used to evaluate any proposed spatial densities
and establish their interpretation.
\subsection{Discrete level contribution}
\label{subsec:discrete_level}
The isoscalar spin-orbit correlation in the nucleon is given by sums over the occupied quark
single-particle levels in the chiral field, including the discrete level and the negative continuum,
see Eqs.~(\ref{spinorbit_kin_sum}) and (\ref{spinorbit_pot_sum}).  We now evaluate the contribution
of the discrete level and discuss its properties.  The contribution of the continuum is evaluated in
Secs.~\ref{subsec:gradient_expansion} and \ref{subsec:numerical} using analytic approximations and
numerical methods.

The quark single-particle Hamitonian Eq.~(\ref{hamiltonian}) commutes with the grand spin operator
$\bm{K} = \bm{J} + \bm{T}$ and the parity operator $\Pi$. The discrete level appears in the $K=0$
and $\Pi= +$ sector of the spectrum. The bound-state wave function has the form
\begin{align}
\Phi_{\rm lev} (\bm{x}) &= \frac{1}{\sqrt{4\pi}}
\left(
\begin{array}{r} f_0 (r) \\[1ex] 
\displaystyle
-i \bm{n}\cdot\bm{\sigma} \, f_1 (r) 
\end{array} \right) \chi ,
\label{level_spinor}
\end{align}
where $\chi$ is the spinor-isospinor wave function where spin 1/2 and isospin 1/2 are coupled to
zero total, $(\bm{\sigma} + \bm{\tau}) \chi = 0$, with $\chi^\dagger \chi = 1$.  The radial wave
functions $f_0$ and $f_1$ obey the equation (in the chiral limit $m=0$)
\begin{align}
& \left(\begin{array}{cc}
M \cos{P(r)} & \displaystyle -\frac{\partial}{\partial r} - \frac{2}{r} + M \sin{P(r)} \\  
\displaystyle \frac{\partial}{\partial r} + M\sin{P(r)} & - M\cos{P(r)}
\end{array}\right)
\nonumber \\
& 
\times \left(\begin{array}{c} f_0(r)  \\[1ex] f_1(r) \end{array}\right) 
= E_{\mathrm{lev}} \left(\begin{array}{c} f_0(r)  \\[1ex] f_1(r) \end{array}\right),
\label{level_radial}
\end{align}
where $P(r)$ is the profile function of the chiral field, Eq.~(\ref{hedgehog}), and
$E_{\mathrm{lev}}$ is the energy eigenvalue. They are normalized as
\begin{align}
\int^{\infty}_{0} dr \, r^{2} \left[f_0^2(r) + f_1^2(r) \right] = 1.
\label{level_normalization}
\end{align}
The subscript on the radial wave functions indicates the orbital angular momentum of the spinor
component.  The upper component has $L = 0$, and the lower component has $L = 1$. For the physical
soliton profile the lower component accounts for $\sim 20\%$ of the normalization integral, showing
that the motion is moderately relativistic. With the variational profile Eq.~(\ref{arctan}) one can
explore the limits of weak binding, where the motion becomes nonrelativistic and the lower component
vanishes, and the limit of strong binding, where the motion is strongly relativistic and upper and
lower components have the same magnitude.

The kinetic part of the spin-orbit correlation resulting from the discrete level is obtained from
Eq.~(\ref{spinorbit_kin_sum}) as
\begin{align}
C^{u+d}_{z}[\textrm{kin, lev}]
&= -\frac{2 N_{c}}{3} \int^{\infty}_{0} dr \, r^2 \, f_1^{2}(r) .
\label{spinorbit_level_kin}
\end{align}
One sees that it arises only from the lower component and represents a relativistic effect.  The
potential part of the spin-orbit correlation resulting from the discrete level is obtained from
Eq.~(\ref{spinorbit_pot_sum}) as
\begin{align}
&C^{u+d}_{z} [\mathrm{pot, lev}] = \frac{ N_{c} M}{3} \int^{\infty}_{0} dr \, r^{3}  \,
\left[ 2 f_0 (r) f_1 (r) \cos{P(r)} \right.
\nonumber \\
& \hspace{6em} \left. - \left(f_0^2 - f_1^{2} \right)\sin{P(r)} \right] .
\label{spinorbit_level_pot}
\end{align}
It involves both the upper and lower components, as well as the classical chiral field.  At the same
time, the vector charge resulting from the discrete level is
\begin{align}
&G^{u+d}_{E}(0)[\textrm{lev}] = N_c \int^{\infty}_{0} dr \, r^{2}
    \left[f_0^{2}(r)+f_1^{2}(r)\right] \; = \; N_c ,
\label{charge_level}
\end{align}
by virtue of the normalization condition Eq.~(\ref{level_normalization}). The discrete level carries
the entire vector charge of the nucleon; the continuum contribution to this quantity sums to zero.
The spin-orbit correlation resulting from the discrete level thus by itself satisfies the sum rule
Eq.~(\ref{sumrule_effective}).  Indeed, using the radial equation Eq.~(\ref{level_radial}) to
rewrite the potential part Eq.~(\ref{spinorbit_level_pot}), and combining it with the kinetic part
Eq.~(\ref{spinorbit_level_kin}), we obtain $-1/2$ times the vector charge of
Eq.~(\ref{charge_level}).

Numerical values of the discrete level contribution to the spin-orbit correlation are computed with
$N_c = 3$ and the physical soliton profile. We obtain
\begin{subequations}
\label{spinorbit_level_numerical}
\begin{align}
C^{u+d}_{z}[\mathrm{kin, lev}] &= -0.41,
\\[1ex]
C^{u+d}_{z}[\mathrm{pot, lev}] &= -1.09,
\\
-\frac{1}{2} G^{u+d}_{E}(0)[\textrm{lev}] &= -\frac{3}{2}. 
\end{align}
\end{subequations}
One observes that the potential term makes a large negative contribution to the spin-orbit
correlation. It has the same sign as the kinetic term and approximately twice the magnitude. The
negative sign of the kinetic term indicates antiparallel alignment of the quark spin and orbital
angular momentum; see Eq.~(\ref{spinorbit_kin_sum}).
\subsection{Gradient expansion}
\label{subsec:gradient_expansion}
We now evaluate the isoscalar spin-orbit correlation in the nucleon using the gradient expansion.
In this method the average of the operator over the quark fields is expanded in gradients of the
classical chiral field of the soliton (see Sec.~\ref{subsec:bosonized}), resulting in analytic
expressions of the nucleon matrix element.  The gradient expansion may be viewed as an approximation
to the total matrix element, including the discrete level and the continuum contributions (see
discussion below).

Applying the gradient expansion to the kinetic part of the isoscalar spin-orbit correlation, using
Eq.~(\ref{tensor_gradient_kin}), we obtain
\begin{align}
&C^{u+d}_z [\mathrm{kin, grad}]
\nonumber \\
&= -\frac{N_{c}}{72\pi^{2}}  \int d^{3}x  \,
\epsilon^{ijk} x^{k} \, \mathrm{tr}_{\textrm{fl}} \left[  \partial_{l} L_{i}L_{j} L_{l} \right], 
\label{spinorbit_gradient_kin}
\end{align}
where $L_i$ is the left current of the chiral field, Eq.~(\ref{left_right_currents}), and the
derivative acts only on the nearest $L$ function [see comment after
Eq.~(\ref{tensor_gradient_kin})].  Applying the same expansion to the potential part, using
Eq.~(\ref{tensor_gradient_pot}), we obtain
\begin{align}
&C^{u+d}_z [\mathrm{pot, grad}] 
\nonumber \\
&= \frac{N_{c}}{48 \pi^{2}} \int d^{3}x \, \epsilon^{ijk} \,
\mathrm{tr}_{\textrm{fl}} [ L_{i}L_{j} L_{k}]
\nonumber \\
&+ \frac{N_{c}}{72\pi^{2}}  \int d^{3}x  \,  \epsilon^{ijk} x^{k} \,
\mathrm{tr}_{\textrm{fl}} \left[  \partial_{l} L_{i} L_{j} L_{l} \right].
\label{spinorbit_gradient_pot}
\end{align}
These expressions should be compared with the well-known result of the gradient expansion of the
nucleon's isovector vector charge
\begin{align}
&G^{u+d}_{E}(0)[\textrm{grad}] = 
-\frac{N_{c}}{24\pi^{2}} \int d^{3} x \, \epsilon^{ijk} \,
\mathrm{tr}_{\textrm{fl}}[ L_{i} L_{j} L_{k}] .
\label{charge_gradient}
\end{align}
This is the winding number of the chiral field (topological charge), which depends only on the
boundary conditions $P(\infty) - P(0)$ but not on the shape of the profile function.  It connects
the baryon number in the quark-based mean-field picture with the winding number in the topological
soliton picture.  One sees that in the kinetic part Eq.~(\ref{spinorbit_gradient_kin}) the function
of the chiral field under the integral is different from the winding number density. The integral is
not topologically invariant (this can be confirmed by evaluating it with different profile functions
satisfying the same boundary conditions). The potential part, Eq.~(\ref{spinorbit_gradient_pot}),
however, has a term that is proportional to the winding number, and another term that cancels the
kinetic term.  Thus the sum of the kinetic and potential terms in the gradient expansion satisfy the
sum rule Eq.~(\ref{sumrule_effective}) with the isoscalar vector charge in the gradient expansion, a
very appealing result. The winding number necessary for this to happen arises only from the
potential term.  This again demonstrates the importance of the potential term of the effective
operator.

It is interesting to evaluate the gradient expansion expressions
Eqs.~(\ref{spinorbit_gradient_kin}), (\ref{spinorbit_gradient_pot}) and (\ref{charge_gradient})
numerically, using the variational profile Eq.~(\ref{arctan}) and $N_c = 3$. We obtain
\begin{subequations}
\label{spinorbit_gradient_numerical}
\begin{align}
C^{u+d}_{z}[\mathrm{kin, grad}] &= -0.67,
\\[1ex]
C^{u+d}_{z}[\mathrm{pot, grad}] &= -0.83,
\\[.5ex]
-\frac{1}{2} G^{u+d}_{E}(0)[\textrm{grad}] &= -\frac{3}{2}. 
\end{align}
\end{subequations}
The results of Eq.~(\ref{spinorbit_gradient_numerical}a) and (b) do not depend on the soliton size
parameter $R$ (this follows from dimensional arguments), but do depend on the specific shape of the
profile function. The isoscalar vector charge in Eq.~(\ref{spinorbit_gradient_numerical}c) does not
depend on either the size or the shape, of course.  One sees that the potential term makes a large
numerical contribution to the spin-orbit correlation in the gradient expansion, of the same sign as
that of the kinetic term.

The gradient expansion represents a formal expansion in the parameter $(MR)^{-1}$, where $M$ is the
dynamical quark mass and $R$ the soliton size [see e.g.\ Eq.~(\ref{arctan})].  Its use as an
approximation to the sums over quark levels for finite $R$ depends on the quantity in question. For
``non-topological'' quantities such as the nucleon mass, isovector axial charge, and others,
successive orders of the gradient expansion are generally nonzero. In this case the expressions at
finite $R$ are used as an approximation to the continuum contribution to the sum over quark levels;
its accuracy has been tested by comparing with exact numerical calculations
\cite{Christov:1995vm}. For the ``topological'' quantity of the the baryon number, the leading
nonzero term of the gradient expansion gives the exact result, while the higher-order terms are zero
\cite{Diakonov:1987ty}. In this case the expressions are finite $R$ are to be regarded as an
approximation to the entire sum over levels, including the discrete level and the continuum.
\subsection{Numerical evaluation}
\label{subsec:numerical}
We now perform an exact numerical evaluation of the isoscalar spin-orbit correlation in the
mean-field picture, including the discrete level and the continuum in the sum over quark levels in
Eqs.~(\ref{spinorbit_kin_sum}) and (\ref{spinorbit_pot_sum}) (see Fig.~\ref{fig:baryon}). The
dynamical quark mass is taken as $M$ = 350 MeV, which is approximately the value obtained from the
instanton vacuum \cite{Diakonov:1985eg,Diakonov:1995qy}.  The soliton profile is the self-consistent
profile obtained by minimizing the energy, Eq.~(\ref{energy_minimum}), with the ultraviolet
regularization implemented through a Pauli-Villars cutoff.  The ultraviolet cutoff is needed only
for the regularization of the soliton energy; the spin-orbit correlation matrix element is
ultraviolet finite and computed without cutoff. The result for the spin-orbit correlation matrix
element is not sensitive to the details of the soliton profile.

%
%
\begin{table}[t]
\centering
\setlength{\tabcolsep}{8pt}
\renewcommand{\arraystretch}{1.2}
\begin{tabular}{c|rrr|r} 
\hline
\hline
 & level  & continuum & total & gradient   \\
\hline
$C^{u+d}_z[\mathrm{kin}]$  & $-0.41$ & $-0.04$ & $-0.45$ & $-0.67$
\\ 
$C^{u+d}_z[\mathrm{pot}]$  & $-1.09$ & $\phantom{-}0.04$ & $-1.05$ & $-0.83$
\\ 
$-\frac{1}{2} G_E^{u+d}(0)$  & $-3/2$ & $ 0 $ & $-3/2$ & $-3/2$
\\ 
\hline 
\hline
\end{tabular}
\caption{Summary of numerical results for the isoscalar quark spin-orbit correlation in the nucleon
in the large-$N_c$ limit. \textit{Rows:} (1)~Spin-orbit correlation from the kinetic term of the
effective operator, $C_z^{u+d}[\textrm{kin}]$, Eq.~(\ref{spinorbit_kin_sum}). (2)~Spin-orbit
correlation from the potential term of the effective operator, $C_z^{u+d}[\textrm{pot}]$,
Eq.~(\ref{spinorbit_pot_sum}). (3)~Isoscalar vector charge $-\frac{1}{2} G_E^{u+d}(0)$,
Eq.~(\ref{charge_sum}).  \textit{Columns:} (a)~Discrete level contribution,
Eq.~(\ref{spinorbit_level_numerical}). (b)~Continuum contribution. (c)~Total result (discrete level
+ continuum). (d)~Gradient expansion, Eq.~(\ref{spinorbit_gradient_numerical}).}
\label{tab:results}
\end{table}
The results are summarized in Table~\ref{tab:results}. One observes: (i)~The continuum contribution
to the spin-orbit correlation matrix elements is numerically small and represents only $\sim 10\%$
of $C_z^{u+d}[\mathrm{kin}]$.  The spin-orbit correlation is dominated by the discrete level
contribution. (ii)~The continuum contributions to $C_z^{u+d}[\mathrm{kin}]$ and
$C_z^{u+d}[\mathrm{pot}]$ have opposite sign and sum to zero.  This is necessary because
$C_z^{u+d}[\mathrm{kin}] + C_z^{u+d}[\mathrm{pot}]$ in the mean-field picture obey the sum rule with
$-\frac{1}{2}G_E^{u + d}(0)$, and the discrete level already accounts for the entire value of the
latter.  (iii)~The potential part of the total spin-orbit correlation (discrete level + continuum)
is numerically large, about twice as large as the kinetic part. This underscores the importance of
the potential part of the effective operator obtained from the instanton vacuum. As demonstrated in
Secs.~\ref{subsec:effective_operators} and \ref{subsec:first_quantized}, the kinetic and potential
terms together are needed to satisfy the sum rule for the spin-orbit correlation.  (iv)~The gradient
expansion provides a reasonable numerical approximation to the exact numerical results for the total
spin-orbit correlation (discrete level + continuum), $-0.67$ vs. $-0.45$ for
$C_z^{u+d}[\mathrm{kin}]$, and $-0.83$ vs.\ $-1.05$ for $C_z^{u+d}[\mathrm{pot}]$.
\section{Discussion}
\label{sec:discussion}
\subsection{Quark model and skyrmion limit}
\label{subsec:quarkmodel_skyrmion}
The mean-field picture of massive quarks moving in a classical chiral field ``interpolates'' between
the nonrelativistic quark model and the chiral soliton (skyrmion) model of the nucleon
\cite{Praszalowicz:1995vi}.  This remarkable property can be demonstrated by parametrizing the
chiral field by a variational profile with spatial size $R$ [see e.g.\ Eq.~(\ref{arctan})] and
studying the behavior of the system in the limits of small and large size, $R \ll M^{-1}$ and
$R \gg M^{-1}$. In the limit $R \ll M^{-1}$ the discrete level becomes weakly bound,
$E_{\rm lev} \ll M$, and the motion of the quarks occupying it becomes non-relativistic.  The
distortion of the continuum levels disappears. In this regime the system exhibits the properties of
the nonrelativistic quark model. In the limit $R \gg M^{-1}$ the discrete level becomes strongly
bound, $E_{\rm lev} < 0$.  The distortion of the continuum levels also becomes strong. In this
regime the properties of the system can be computed by the gradient expansion in the chiral field,
and the system has the properties of a soliton of the chiral Lagrangian.  It is instructive to
compute baryon observables as functions of the size parameter $R$ and study their behavior in the
two limits.

%
%
\begin{figure}[t]
\centering
\includegraphics[width=1.\columnwidth]{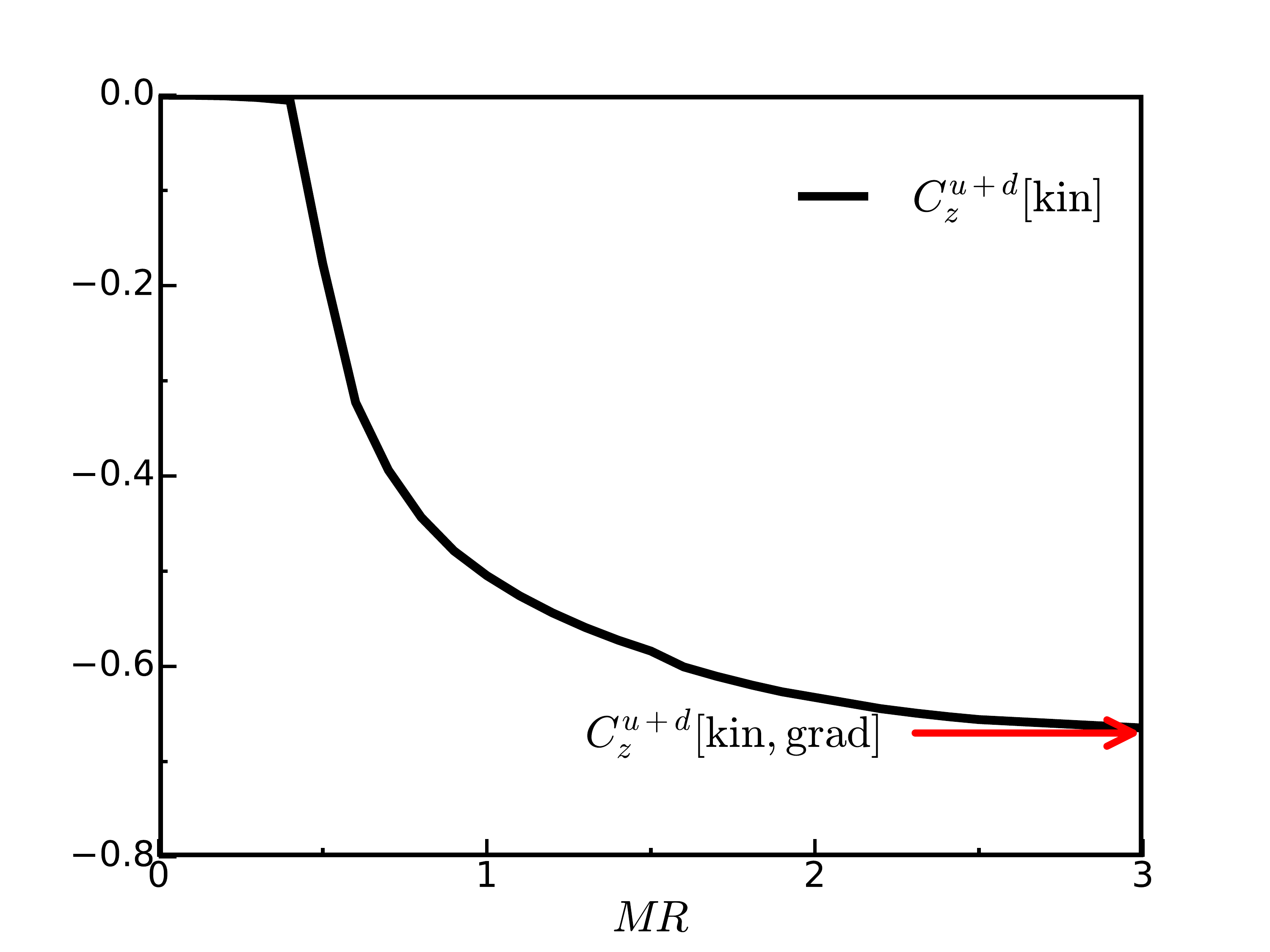}
\caption{The isoscalar spin-orbit correlation $C_z^{u+d}[\mathrm{kin}]$ computed with the
variational profile function Eq.~(\ref{arctan}), as a function of the dimensionless size parameter
$MR$.  The arrow represents the gradient expansion result $C_z^{u+d}[\mathrm{kin, grad}] = -0.67$,
Eq.~(\ref{spinorbit_gradient_numerical}a).}
\label{fig_sizedep}
\end{figure}
Figure~\ref{fig_sizedep} shows the isoscalar spin-orbit correlation
$C_z^{u+d}[\mathrm{kin}]$, evaluated with the variational profile Eq.~(\ref{arctan}), as a function
of the dimensionless size parameter $MR$.  (The plot shows only the kinetic part
$C_z^{u+d}[\mathrm{kin}]$; the potential part $C_z^{u+d}[\mathrm{pot}]$ can be inferred from the sum
rule $C_z^{u+d}[\mathrm{kin}] + C_z^{u+d}[\mathrm{pot}] = -3/2$, which is satisfied at all values of
$MR$.) The physical mean-field solution corresponds approximately to $MR \approx 1$
\cite{Diakonov:1985eg}.  One observes:

(i)~For decreasing size $MR < 1$, $C_z^{u+d}[\mathrm{kin}]$ strongly decreases in magnitude. This
happens because the discrete level contribution is proportional to the lower component of the
bound-state wave function, Eq.~(\ref{level_spinor}), which vanishes in the nonrelativistic limit,
\begin{align}
|f_1(r)| \rightarrow 0.
\end{align}

(ii)~At $MR < 0.4$ the chiral field is no longer strong enough to create a quark bound state, and
the discrete level disappears.  This regime does not correspond to physical baryons; it is shown
only to illustrate what happens to the system if we study it as a function of the size of the chiral
field.

(iii)~For increasing size $MR > 1$, $C_z^{u+d}[\mathrm{kin}]$ increases in magnitude. This happens
because the relativistic effects in the discrete level become stronger, and the contribution of the
continuum levels becomes non-negligible.

(iv)~At $MR \gg 1$, $C_z^{u+d}[\mathrm{kin}]$ can be compared with the result of the gradient
expansion (see Sec.~\ref{subsec:gradient_expansion}). The gradient expansion predicts the size
dependence as constant, \begin{align} C^{u+d}_{z}[\mathrm{kin}] \sim (MR)^0 , \end{align} as can be
inferred from the powers of coordinates and derivatives of the chiral field in the integral of
Eq.~(\ref{spinorbit_gradient_kin}). The numerical results in Figure~\ref{fig_sizedep} indeed
approach such a dependence. They also reproduce the predicted asymptotic value of $-0.67$,
Eq.~(\ref{spinorbit_gradient_numerical}).

Some comments are in order regarding the role of the sign of the energy of the discrete level.
$E_{\rm lev}$ crosses zero at $MR \sim 1.5$ and becomes negative for larger radii.  For physical
quantities represented by sums over occupied levels (discrete + continuum) there is nothing special
about the discrete level crossing zero, as they depend continuously on the parameter $R$. Indeed,
Fig.~\ref{fig_sizedep} shows that $C_z^{u+d}[\mathrm{kin}]$ is continuous at $MR \sim 1.5$.
However, the interpretation of some quantities changes when the discrete level energy becomes
negative.  The identification of the baryon number as the topological winding number of the chiral
field emerges in the region where the discrete level energy is negative.

The dependence on the size of the chiral field elucidates the role of relativistic and interaction
effects in the spin-orbit correlations. Similar studies have been performed for the nucleon's
isoscalar axial charge and the tensor charges \cite{Praszalowicz:1995vi,Kim:1995bq}.
\subsection{$t$-dependent form factors}  
%
%
\begin{figure}[t]
\centering
\includegraphics[width=1.\columnwidth]{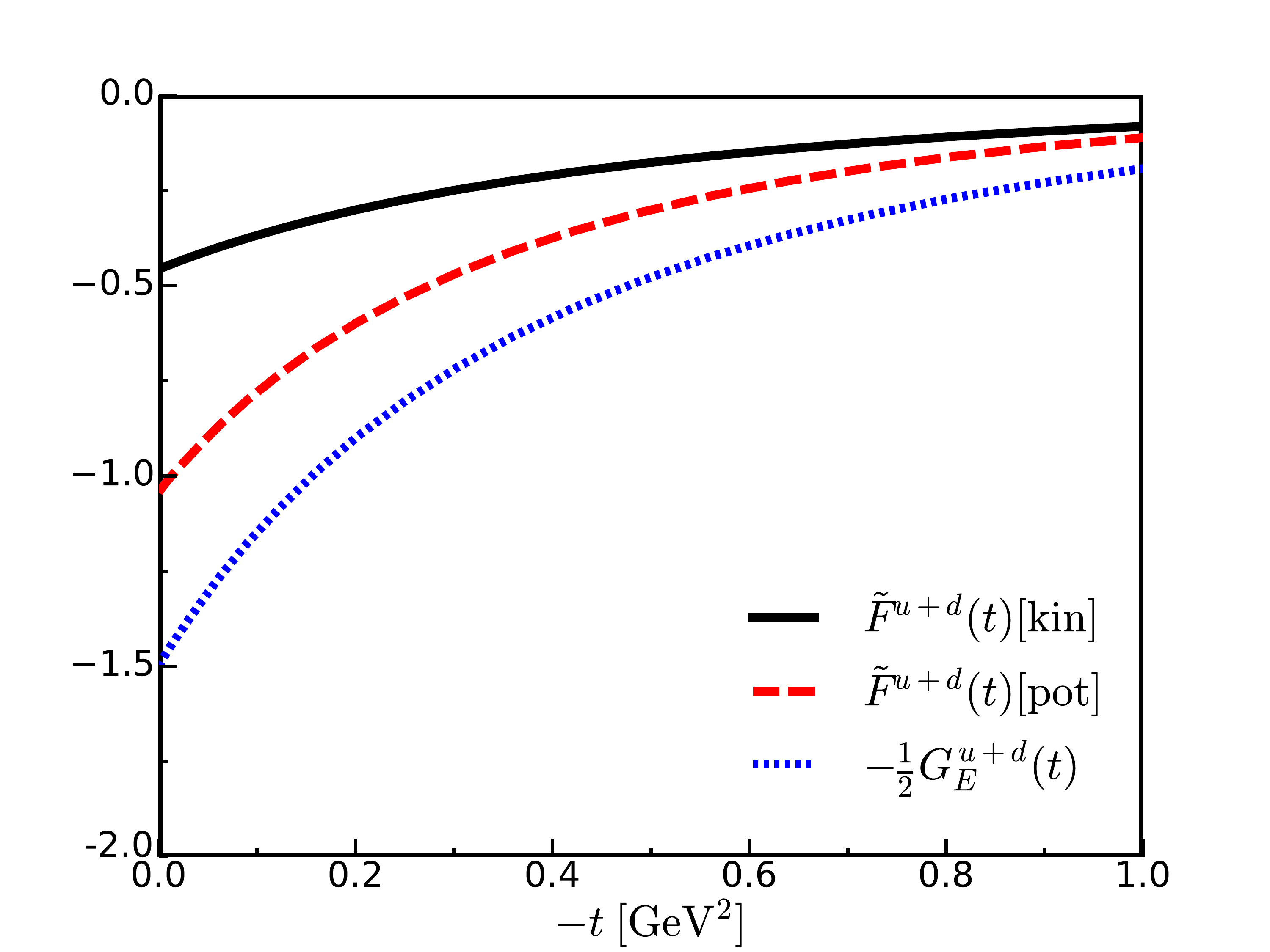}
\caption{The $t$-dependence of the form factor $\tilde{F}^{u+d}(t)$ of the parity-odd tensor
operator, Eq.~(\ref{parametrization}), and of the vector form factor $G_E^{u+d}(t)$, as obtained in
the large-$N_c$ mean-field picture. \textit{Solid line:} Kinetic part
$\tilde{F}^{u+d}(t)[\textrm{kin}]$.  \textit{Dashed line:} Potential part
$\tilde{F}^{u+d}(t)[\textrm{pot}]$.  \textit{Dotted line:} $-\frac{1}{2} G_E^{u+d}(t) =
\tilde{F}^{u+d}(t)[\textrm{kin}] + \tilde{F}^{u+d}(t)[\textrm{pot}]$.}
\label{fig:formfactors}
\end{figure}
The spin-orbit correlations are obtained from the invariant form factors of the tensor operator
Eq.~(\ref{operator_def}) at $t = 0$. Further information on nucleon structure is contained in the
$t$-dependence of the form factors. Using the effective operators from the instanton vacuum and the
mean-field picture of the nucleon, we can compute the invariant form factors at finite momentum
transfers $|t| = \mathcal{O}(N_c^0)$.

Figure~\ref{fig:formfactors} shows the results for the $t$-dependent form factors
$\tilde{F}^{u+d}(t) = \frac{1}{2} \tilde{C}^{u+d}(t)$ (up to $1/N_c$ corrections), whose values at
$t=0$ give $C_z^{u+d}$, and the $t$-dependent isoscalar vector form factor $G_E^{u+d}(t)$. The plot
shows the individual contributions of the kinetic and potential terms of the effective operators to
the form factors.  One observes: (i)~Both $\tilde{F}^{u+d}(t)[\textrm{kin}]$ and
$\tilde{F}^{u+d}(t)[\textrm{pot}]$ decrease as functions of $|t|$. (ii)~The $t$-dependence of
$\tilde{F}^{u+d}(t)[\textrm{pot}]$ is similar to that of $G_E^{u+d}(t)$; the $t$-dependence of
$\tilde{F}^{u+d}(t)[\textrm{kin}]$ is slower than that of $G_E^{u+d}(t)$, indicating a larger
effective mass scale in the kinetic form factor.

The physical interpretation of the form factors of the parity-odd tensor operator
Eq.~(\ref{operator_def}) (densities, radii) and their connection with the mechanical properties of
the nucleon are an interesting problem for further study.
\subsection{$N$-$\Delta$ transitions}
\label{subsec:ndelta}
The isovector tensor operator Eq.~(\ref{operator_isospin}) also has matrix elements for transitions
between $N$ and $\Delta$ states. Their large-$N_c$ predictions can be obtained from
Eqs.~(\ref{matrixelement_largenc}b), (d), and (f) by evaluating the expressions with the
corresponding $N$-$\Delta$ transition matrix elements of the spin-flavor operators
($S = T = 1/2, S' = T' = 3/2$),
\begin{subequations}
\begin{align}
\sqrt{\frac{2}{3}} \langle D^{3i} \rangle_{B'B}
&= -C^{\frac{3}{2} T_{3}'}_{\frac{1}{2} T_{3} 1 0 } \sqrt{2} \tilde{S}^i_{S_3'S_3} ,
\\[1ex]
\sqrt{\frac{2}{3}} \langle \{ S^i, D^{3j} \} \rangle_{B'B}
&= -C^{\frac{3}{2} T_{3}'}_{\frac{1}{2} T_{3} 1 0 }
\nonumber \\
&\times \sqrt{2} \left[ \tilde{Q}^{ij}_{S_3'S_3}
+ i \epsilon^{ijk} \tilde{S}^k_{S_3'S_3} \right] ,
\end{align}
\end{subequations}
where the $N$-$\Delta$ transition matrix elements of the multipole operators are defined as
\begin{subequations}
\begin{align}
& \tilde{S}^{i}_{\sigma' \sigma} \equiv \frac{1}{2} \sqrt{\frac{3}{2}} \sum_{\lambda'}
C^{\frac32 \sigma' }_{1 \lambda' \frac12 \sigma} \epsilon^{*i}_{\lambda'},
\\
& \tilde{Q}^{ij}_{\sigma' \sigma} \equiv \sqrt{\frac{3}{2}} \sum_{\lambda' s'}
C^{\frac32 \sigma' }_{1 \lambda' \frac12 s'} S^{i}_{s' \sigma} \epsilon^{*j}_{\lambda'}.
\end{align}
\end{subequations}
Here $\epsilon^{*i}_{\lambda'}$ are the spin-1 polarization vectors, which can be understood as the
transformation matrix from spherical to Cartesian components in the spin-1 case.
$S^{i}_{s'\sigma} = [\sigma^{i}/2]_{s' \sigma}$ is defined through the matrix elements of the Pauli
matrices.

The interpretation of the $N$-$\Delta$ transition matrix elements in terms of spin-orbit
correlations in QCD is still to be developed. We note that the $1/N_c$ expansion of the light-front
components of the transition matrix element of the tensor operator can be performed using the
``matching'' procedure of Ref.~\cite{Kim:2023xvw} and does not require an invariant form factor
decomposition of the $N$-$\Delta$ transition matrix element.
\section{Summary and extensions}
\label{sec:summary}
In this article we have studied the QCD spin-orbit correlations on the basis of the effective
dynamics emerging from the spontaneous breaking of chiral symmetry. The effective operators
representing the QCD operators are derived using the microscopic model of the chiral-symmetry
breaking gauge fields provided by the instanton vacuum. The effective operators contain kinetic and
potential terms and satisfy the same equation-of-motion relations as the QCD operators.  The nucleon
is obtained as a mean-field solution of the effective dynamics describing quarks moving in a
classical chiral field. The nucleon matrix elements are evaluated in the $1/N_c$ expansion and
receive contributions from the discrete and continuum levels in the quark single-particle spectrum.
The main results can be summarized as follows:
\begin{itemize}
\item[(i)] The isoscalar spin-orbit correlation $C^{u+d}_{z}$ is $\mathcal{O}(N_c)$ and leading in
the $1/N_c$ expansion. It can be extracted from the form factors $\tilde{F}^{u+d}(t)$ or
$\frac{1}{2} \tilde{C}^{u+d}(t)$ of the parity-odd tensor operator, which coincide in leading order
of $1/N_c$.  The isovector spin-orbit correlation $C^{u-d}_{z}$ is $\mathcal{O}(N_c^0)$ and
suppressed.
\item[(ii)] The nucleon matrix elements of the effective operators in the effective theory satisfy
the equation-of-motion relations thanks to the presence of kinetic and potential terms in the
effective operators.  For the isoscalar spin-orbit correlation, the equation-of-motion relation is
satisfied independently in each quark single-particle level in the nucleon.
\item[(iii)] In the first-quantized representation in quark single-particle levels, the kinetic term
of the effective operator takes the form $\bm{L}\cdot\bm{\Sigma}$ and describes the correlation of
the quark spin and orbital angular momentum. The potential term of the effective operator takes the
form $i\bm{x}\cdot\bm{\Sigma} \, \gamma^0\gamma^5 M U^{\gamma_5}(\bm{x})$ and describes a coupling
of the quarks to the classical chiral field of the nucleon. This form of spin-orbit correlation
become possible because the chiral field provides a local pseudovector that can couple to the quark
spin. The presence of this structure qualitatively changes the picture of spin-orbit correlations
compared to simple quark model expectations.
\item[(iv)] The numerical values of the matrix elements are obtained as
$C^{u+d}_{z}[\mathrm{kin}] = -0.41$ and $C^{u+d}_{z}[\mathrm{pot}] = -1.09$. The negative value of
the kinetic part indicates that the quark spin and orbital angular momentum in the nucleon are
anti-aligned in the flavor-singlet sector.  The value of the potential part is $\sim 2\times$ larger
in magnitude and shows the essential role of interaction effects in the overall spin-orbit
correlation matrix element.
\item[(v)] The gradient expansion of the isoscalar spin-orbit correlation matrix elements reproduces
the isoscalar vector charge (baryon number) as given by the winding number of the soliton field.
The topological current arises entirely from the potential term of the spin-orbit operator.
\end{itemize}
In the present study we have evaluated the isoscalar (flavor-singlet) spin-orbit correlation in the
nucleon, which is leading in the $1/N_c$ expansion. The framework described here can also be used to
evaluate the isovector (flavor-nonsinglet) spin-orbit correlation. In this case the zero mode
quantization procedure must take into account the $\mathcal{O}(N_c^{-1})$ finite velocities of the
collective motion of the mean field, and the nucleon matrix elements are obtained as double sums
over quark single-particle levels.

The framework can also be used to compute the $N$-$\Delta$ transition matrix elements of the
isovector parity-odd tensor operator Eq.~(\ref{operator_def}). The interpretation of QCD spin-orbit
correlations in $N$-$\Delta$ transitions can be developed along the same lines as that of the total
transition angular momentum, using the formulation in terms of light-front transition densities
\cite{Kim:2023xvw}.  The $1/N_c$ expansion of the light-front components of the transition matrix
elements of the tensor operator can be performed efficiently, using the procedure of matching them
with 3-dimensional multipoles in a special symmetric frame \cite{Kim:2023xvw}.

The effective operator method developed in Refs.~\cite{Diakonov:1995qy,Balla:1997hf,Kim:2023pll} can
be extended to compute nucleon matrix elements of several other QCD operators of interest. One
example is the flavor decomposition of the quark angular momentum measured by the the twist-3 part
of the quark EMT. Another example are the higher moments of the twist-3 quark GPDs appearing in the
factorization of hard exclusive electroproduction processes.

\section*{Acknowledgments}
This material is based upon work supported by the U.S.~Department of Energy, Office of Science,
Office of Nuclear Physics under contract DE-AC05-06OR23177.
This work is also supported by the Basic Science Research Program through the National Research
Foundation of Korea funded by the Korean government (Ministry of Education, Science and Technology,
MEST), Grant-No. 2021R1A2C2093368 and 2018R1A5A1025563 (HChK).
It is also supported by the France Excellence scholarship through Campus France funded by the French
government (Minist\`ere de l’Europe et des Affaires \'etrang\`eres), Grant-No.  141295X (HYW).

The research reported here takes place in the context of the Topical Collaboration ``3D quark-gluon
structure of hadrons: mass, spin, tomography'' (Quark-Gluon Tomography Collaboration) supported by
the U.S.~Department of Energy, Office of Science, Office of Nuclear Physics.
\bibliography{spinorbit}

\begin{thebibliography}{45}%
\makeatletter
\providecommand \@ifxundefined [1]{%
 \@ifx{#1\undefined}
}%
\providecommand \@ifnum [1]{%
 \ifnum #1\expandafter \@firstoftwo
 \else \expandafter \@secondoftwo
 \fi
}%
\providecommand \@ifx [1]{%
 \ifx #1\expandafter \@firstoftwo
 \else \expandafter \@secondoftwo
 \fi
}%
\providecommand \natexlab [1]{#1}%
\providecommand \enquote  [1]{``#1''}%
\providecommand \bibnamefont  [1]{#1}%
\providecommand \bibfnamefont [1]{#1}%
\providecommand \citenamefont [1]{#1}%
\providecommand \href@noop [0]{\@secondoftwo}%
\providecommand \href [0]{\begingroup \@sanitize@url \@href}%
\providecommand \@href[1]{\@@startlink{#1}\@@href}%
\providecommand \@@href[1]{\endgroup#1\@@endlink}%
\providecommand \@sanitize@url [0]{\catcode `\\12\catcode `\$12\catcode
  `\&12\catcode `\#12\catcode `\^12\catcode `\_12\catcode `\%12\relax}%
\providecommand \@@startlink[1]{}%
\providecommand \@@endlink[0]{}%
\providecommand \url  [0]{\begingroup\@sanitize@url \@url }%
\providecommand \@url [1]{\endgroup\@href {#1}{\urlprefix }}%
\providecommand \urlprefix  [0]{URL }%
\providecommand \Eprint [0]{\href }%
\providecommand \doibase [0]{https://doi.org/}%
\providecommand \selectlanguage [0]{\@gobble}%
\providecommand \bibinfo  [0]{\@secondoftwo}%
\providecommand \bibfield  [0]{\@secondoftwo}%
\providecommand \translation [1]{[#1]}%
\providecommand \BibitemOpen [0]{}%
\providecommand \bibitemStop [0]{}%
\providecommand \bibitemNoStop [0]{.\EOS\space}%
\providecommand \EOS [0]{\spacefactor3000\relax}%
\providecommand \BibitemShut  [1]{\csname bibitem#1\endcsname}%
\let\auto@bib@innerbib\@empty
\bibitem [{\citenamefont {Aidala}\ \emph {et~al.}(2013)\citenamefont {Aidala},
  \citenamefont {Bass}, \citenamefont {Hasch},\ and\ \citenamefont
  {Mallot}}]{Aidala:2012mv}%
  \BibitemOpen
  \bibfield  {author} {\bibinfo {author} {\bibfnamefont {C.~A.}\ \bibnamefont
  {Aidala}}, \bibinfo {author} {\bibfnamefont {S.~D.}\ \bibnamefont {Bass}},
  \bibinfo {author} {\bibfnamefont {D.}~\bibnamefont {Hasch}},\ and\ \bibinfo
  {author} {\bibfnamefont {G.~K.}\ \bibnamefont {Mallot}},\ }\bibfield  {title}
  {\bibinfo {title} {{The Spin Structure of the Nucleon}},\ }\href
  {https://doi.org/10.1103/RevModPhys.85.655} {\bibfield  {journal} {\bibinfo
  {journal} {Rev. Mod. Phys.}\ }\textbf {\bibinfo {volume} {85}},\ \bibinfo
  {pages} {655} (\bibinfo {year} {2013})},\ \Eprint
  {https://arxiv.org/abs/1209.2803} {arXiv:1209.2803 [hep-ph]} \BibitemShut
  {NoStop}%
\bibitem [{\citenamefont {Deur}\ \emph {et~al.}(2019)\citenamefont {Deur},
  \citenamefont {Brodsky},\ and\ \citenamefont {De~T\'eramond}}]{Deur:2018roz}%
  \BibitemOpen
  \bibfield  {author} {\bibinfo {author} {\bibfnamefont {A.}~\bibnamefont
  {Deur}}, \bibinfo {author} {\bibfnamefont {S.~J.}\ \bibnamefont {Brodsky}},\
  and\ \bibinfo {author} {\bibfnamefont {G.~F.}\ \bibnamefont
  {De~T\'eramond}},\ }\bibfield  {title} {\bibinfo {title} {{The Spin Structure
  of the Nucleon}},\ }\href {https://doi.org/10.1088/1361-6633/ab0b8f}
  {\bibfield  {journal} {\bibinfo  {journal} {Rept. Prog. Phys.}\ }\textbf
  {\bibinfo {volume} {82}},\ \bibinfo {pages} {076201} (\bibinfo {year}
  {2019})},\ \Eprint {https://arxiv.org/abs/1807.05250} {arXiv:1807.05250
  [hep-ph]} \BibitemShut {NoStop}%
\bibitem [{\citenamefont {Ji}(1997)}]{Ji:1996ek}%
  \BibitemOpen
  \bibfield  {author} {\bibinfo {author} {\bibfnamefont {X.-D.}\ \bibnamefont
  {Ji}},\ }\bibfield  {title} {\bibinfo {title} {{Gauge-Invariant Decomposition
  of Nucleon Spin}},\ }\href {https://doi.org/10.1103/PhysRevLett.78.610}
  {\bibfield  {journal} {\bibinfo  {journal} {Phys. Rev. Lett.}\ }\textbf
  {\bibinfo {volume} {78}},\ \bibinfo {pages} {610} (\bibinfo {year} {1997})},\
  \Eprint {https://arxiv.org/abs/hep-ph/9603249} {arXiv:hep-ph/9603249}
  \BibitemShut {NoStop}%
\bibitem [{\citenamefont {Goeke}\ \emph {et~al.}(2001)\citenamefont {Goeke},
  \citenamefont {Polyakov},\ and\ \citenamefont
  {Vanderhaeghen}}]{Goeke:2001tz}%
  \BibitemOpen
  \bibfield  {author} {\bibinfo {author} {\bibfnamefont {K.}~\bibnamefont
  {Goeke}}, \bibinfo {author} {\bibfnamefont {M.~V.}\ \bibnamefont
  {Polyakov}},\ and\ \bibinfo {author} {\bibfnamefont {M.}~\bibnamefont
  {Vanderhaeghen}},\ }\bibfield  {title} {\bibinfo {title} {{Hard exclusive
  reactions and the structure of hadrons}},\ }\href
  {https://doi.org/10.1016/S0146-6410(01)00158-2} {\bibfield  {journal}
  {\bibinfo  {journal} {Prog. Part. Nucl. Phys.}\ }\textbf {\bibinfo {volume}
  {47}},\ \bibinfo {pages} {401} (\bibinfo {year} {2001})},\ \Eprint
  {https://arxiv.org/abs/hep-ph/0106012} {arXiv:hep-ph/0106012} \BibitemShut
  {NoStop}%
\bibitem [{\citenamefont {Diehl}(2003)}]{Diehl:2003ny}%
  \BibitemOpen
  \bibfield  {author} {\bibinfo {author} {\bibfnamefont {M.}~\bibnamefont
  {Diehl}},\ }\bibfield  {title} {\bibinfo {title} {{Generalized parton
  distributions}},\ }\href {https://doi.org/10.1016/j.physrep.2003.08.002}
  {\bibfield  {journal} {\bibinfo  {journal} {Phys. Rept.}\ }\textbf {\bibinfo
  {volume} {388}},\ \bibinfo {pages} {41} (\bibinfo {year} {2003})},\ \Eprint
  {https://arxiv.org/abs/hep-ph/0307382} {arXiv:hep-ph/0307382} \BibitemShut
  {NoStop}%
\bibitem [{\citenamefont {Belitsky}\ and\ \citenamefont
  {Radyushkin}(2005)}]{Belitsky:2005qn}%
  \BibitemOpen
  \bibfield  {author} {\bibinfo {author} {\bibfnamefont {A.~V.}\ \bibnamefont
  {Belitsky}}\ and\ \bibinfo {author} {\bibfnamefont {A.~V.}\ \bibnamefont
  {Radyushkin}},\ }\bibfield  {title} {\bibinfo {title} {{Unraveling hadron
  structure with generalized parton distributions}},\ }\href
  {https://doi.org/10.1016/j.physrep.2005.06.002} {\bibfield  {journal}
  {\bibinfo  {journal} {Phys. Rept.}\ }\textbf {\bibinfo {volume} {418}},\
  \bibinfo {pages} {1} (\bibinfo {year} {2005})},\ \Eprint
  {https://arxiv.org/abs/hep-ph/0504030} {arXiv:hep-ph/0504030} \BibitemShut
  {NoStop}%
\bibitem [{\citenamefont {Leader}\ and\ \citenamefont
  {Lorc\'e}(2014)}]{Leader:2013jra}%
  \BibitemOpen
  \bibfield  {author} {\bibinfo {author} {\bibfnamefont {E.}~\bibnamefont
  {Leader}}\ and\ \bibinfo {author} {\bibfnamefont {C.}~\bibnamefont
  {Lorc\'e}},\ }\bibfield  {title} {\bibinfo {title} {{The angular momentum
  controversy: What\textquoteright{}s it all about and does it matter?}},\
  }\href {https://doi.org/10.1016/j.physrep.2014.02.010} {\bibfield  {journal}
  {\bibinfo  {journal} {Phys. Rept.}\ }\textbf {\bibinfo {volume} {541}},\
  \bibinfo {pages} {163} (\bibinfo {year} {2014})},\ \Eprint
  {https://arxiv.org/abs/1309.4235} {arXiv:1309.4235 [hep-ph]} \BibitemShut
  {NoStop}%
\bibitem [{\citenamefont {Lorc\'e}(2014)}]{Lorce:2014mxa}%
  \BibitemOpen
  \bibfield  {author} {\bibinfo {author} {\bibfnamefont {C.}~\bibnamefont
  {Lorc\'e}},\ }\bibfield  {title} {\bibinfo {title} {{Spin\textendash{}orbit
  correlations in the nucleon}},\ }\href
  {https://doi.org/10.1016/j.physletb.2014.06.068} {\bibfield  {journal}
  {\bibinfo  {journal} {Phys. Lett. B}\ }\textbf {\bibinfo {volume} {735}},\
  \bibinfo {pages} {344} (\bibinfo {year} {2014})},\ \Eprint
  {https://arxiv.org/abs/1401.7784} {arXiv:1401.7784 [hep-ph]} \BibitemShut
  {NoStop}%
\bibitem [{\citenamefont {Diakonov}(2003)}]{Diakonov:2002fq}%
  \BibitemOpen
  \bibfield  {author} {\bibinfo {author} {\bibfnamefont {D.}~\bibnamefont
  {Diakonov}},\ }\bibfield  {title} {\bibinfo {title} {{Instantons at work}},\
  }\href {https://doi.org/10.1016/S0146-6410(03)90014-7} {\bibfield  {journal}
  {\bibinfo  {journal} {Prog. Part. Nucl. Phys.}\ }\textbf {\bibinfo {volume}
  {51}},\ \bibinfo {pages} {173} (\bibinfo {year} {2003})},\ \Eprint
  {https://arxiv.org/abs/hep-ph/0212026} {arXiv:hep-ph/0212026} \BibitemShut
  {NoStop}%
\bibitem [{\citenamefont {Sch\"afer}\ and\ \citenamefont
  {Shuryak}(1998)}]{Schafer:1996wv}%
  \BibitemOpen
  \bibfield  {author} {\bibinfo {author} {\bibfnamefont {T.}~\bibnamefont
  {Sch\"afer}}\ and\ \bibinfo {author} {\bibfnamefont {E.~V.}\ \bibnamefont
  {Shuryak}},\ }\bibfield  {title} {\bibinfo {title} {{Instantons in QCD}},\
  }\href {https://doi.org/10.1103/RevModPhys.70.323} {\bibfield  {journal}
  {\bibinfo  {journal} {Rev. Mod. Phys.}\ }\textbf {\bibinfo {volume} {70}},\
  \bibinfo {pages} {323} (\bibinfo {year} {1998})},\ \Eprint
  {https://arxiv.org/abs/hep-ph/9610451} {arXiv:hep-ph/9610451} \BibitemShut
  {NoStop}%
\bibitem [{\citenamefont {Shuryak}(1982{\natexlab{a}})}]{Shuryak:1981ff}%
  \BibitemOpen
  \bibfield  {author} {\bibinfo {author} {\bibfnamefont {E.~V.}\ \bibnamefont
  {Shuryak}},\ }\bibfield  {title} {\bibinfo {title} {{The Role of Instantons
  in Quantum Chromodynamics. 1. Physical Vacuum}},\ }\href
  {https://doi.org/10.1016/0550-3213(82)90478-3} {\bibfield  {journal}
  {\bibinfo  {journal} {Nucl. Phys. B}\ }\textbf {\bibinfo {volume} {203}},\
  \bibinfo {pages} {93} (\bibinfo {year} {1982}{\natexlab{a}})}\BibitemShut
  {NoStop}%
\bibitem [{\citenamefont {Shuryak}(1982{\natexlab{b}})}]{Shuryak:1982dp}%
  \BibitemOpen
  \bibfield  {author} {\bibinfo {author} {\bibfnamefont {E.~V.}\ \bibnamefont
  {Shuryak}},\ }\bibfield  {title} {\bibinfo {title} {{The Role of Instantons
  in Quantum Chromodynamics. 2. Hadronic Structure}},\ }\href
  {https://doi.org/10.1016/0550-3213(82)90479-5} {\bibfield  {journal}
  {\bibinfo  {journal} {Nucl. Phys. B}\ }\textbf {\bibinfo {volume} {203}},\
  \bibinfo {pages} {116} (\bibinfo {year} {1982}{\natexlab{b}})}\BibitemShut
  {NoStop}%
\bibitem [{\citenamefont {Diakonov}\ and\ \citenamefont
  {Petrov}(1984)}]{Diakonov:1983hh}%
  \BibitemOpen
  \bibfield  {author} {\bibinfo {author} {\bibfnamefont {D.}~\bibnamefont
  {Diakonov}}\ and\ \bibinfo {author} {\bibfnamefont {V.~Y.}\ \bibnamefont
  {Petrov}},\ }\bibfield  {title} {\bibinfo {title} {{Instanton Based Vacuum
  from Feynman Variational Principle}},\ }\href
  {https://doi.org/10.1016/0550-3213(84)90432-2} {\bibfield  {journal}
  {\bibinfo  {journal} {Nucl. Phys. B}\ }\textbf {\bibinfo {volume} {245}},\
  \bibinfo {pages} {259} (\bibinfo {year} {1984})}\BibitemShut {NoStop}%
\bibitem [{\citenamefont {Diakonov}\ and\ \citenamefont
  {Petrov}(1986)}]{Diakonov:1985eg}%
  \BibitemOpen
  \bibfield  {author} {\bibinfo {author} {\bibfnamefont {D.}~\bibnamefont
  {Diakonov}}\ and\ \bibinfo {author} {\bibfnamefont {V.~Y.}\ \bibnamefont
  {Petrov}},\ }\bibfield  {title} {\bibinfo {title} {{A Theory of Light Quarks
  in the Instanton Vacuum}},\ }\href
  {https://doi.org/10.1016/0550-3213(86)90011-8} {\bibfield  {journal}
  {\bibinfo  {journal} {Nucl. Phys. B}\ }\textbf {\bibinfo {volume} {272}},\
  \bibinfo {pages} {457} (\bibinfo {year} {1986})}\BibitemShut {NoStop}%
\bibitem [{\citenamefont {Diakonov}\ and\ \citenamefont
  {Petrov}()}]{Diakonov:1986aj}%
  \BibitemOpen
  \bibfield  {author} {\bibinfo {author} {\bibfnamefont {D.}~\bibnamefont
  {Diakonov}}\ and\ \bibinfo {author} {\bibfnamefont {V.~Y.}\ \bibnamefont
  {Petrov}},\ }\href@noop {} {\bibinfo {title} {{Spontaneous breaking of chiral
  symmetry in the instanton vacuum}}},\ \bibinfo {howpublished}
  {LENINGRAD-86-1153. Published (in Russian) in: Hadron Matter under Extreme
  Conditions, Eds.\ G.~M.~Zinovev and V.~P.~Shelest, Naukova Dumka, Kiev
  (1986), p. 192.}\BibitemShut {Stop}%
\bibitem [{\citenamefont {Diakonov}\ \emph
  {et~al.}(1996{\natexlab{a}})\citenamefont {Diakonov}, \citenamefont
  {Polyakov},\ and\ \citenamefont {Weiss}}]{Diakonov:1995qy}%
  \BibitemOpen
  \bibfield  {author} {\bibinfo {author} {\bibfnamefont {D.}~\bibnamefont
  {Diakonov}}, \bibinfo {author} {\bibfnamefont {M.~V.}\ \bibnamefont
  {Polyakov}},\ and\ \bibinfo {author} {\bibfnamefont {C.}~\bibnamefont
  {Weiss}},\ }\bibfield  {title} {\bibinfo {title} {{Hadronic matrix elements
  of gluon operators in the instanton vacuum}},\ }\href
  {https://doi.org/10.1016/0550-3213(95)00675-3} {\bibfield  {journal}
  {\bibinfo  {journal} {Nucl. Phys. B}\ }\textbf {\bibinfo {volume} {461}},\
  \bibinfo {pages} {539} (\bibinfo {year} {1996}{\natexlab{a}})},\ \Eprint
  {https://arxiv.org/abs/hep-ph/9510232} {arXiv:hep-ph/9510232} \BibitemShut
  {NoStop}%
\bibitem [{\citenamefont {Kacir}\ \emph {et~al.}(1999)\citenamefont {Kacir},
  \citenamefont {Prakash},\ and\ \citenamefont {Zahed}}]{Kacir:1996qn}%
  \BibitemOpen
  \bibfield  {author} {\bibinfo {author} {\bibfnamefont {M.}~\bibnamefont
  {Kacir}}, \bibinfo {author} {\bibfnamefont {M.}~\bibnamefont {Prakash}},\
  and\ \bibinfo {author} {\bibfnamefont {I.}~\bibnamefont {Zahed}},\ }\bibfield
   {title} {\bibinfo {title} {{Hadrons and QCD instantons: A Bosonized view}},\
  }\href@noop {} {\bibfield  {journal} {\bibinfo  {journal} {Acta Phys. Polon.
  B}\ }\textbf {\bibinfo {volume} {30}},\ \bibinfo {pages} {287} (\bibinfo
  {year} {1999})},\ \Eprint {https://arxiv.org/abs/hep-ph/9602314}
  {arXiv:hep-ph/9602314} \BibitemShut {NoStop}%
\bibitem [{\citenamefont {Diakonov}\ \emph {et~al.}(1988)\citenamefont
  {Diakonov}, \citenamefont {Petrov},\ and\ \citenamefont
  {Pobylitsa}}]{Diakonov:1987ty}%
  \BibitemOpen
  \bibfield  {author} {\bibinfo {author} {\bibfnamefont {D.}~\bibnamefont
  {Diakonov}}, \bibinfo {author} {\bibfnamefont {V.~Y.}\ \bibnamefont
  {Petrov}},\ and\ \bibinfo {author} {\bibfnamefont {P.~V.}\ \bibnamefont
  {Pobylitsa}},\ }\bibfield  {title} {\bibinfo {title} {{A Chiral Theory of
  Nucleons}},\ }\href {https://doi.org/10.1016/0550-3213(88)90443-9} {\bibfield
   {journal} {\bibinfo  {journal} {Nucl. Phys. B}\ }\textbf {\bibinfo {volume}
  {306}},\ \bibinfo {pages} {809} (\bibinfo {year} {1988})}\BibitemShut
  {NoStop}%
\bibitem [{\citenamefont {Witten}(1979)}]{Witten:1979kh}%
  \BibitemOpen
  \bibfield  {author} {\bibinfo {author} {\bibfnamefont {E.}~\bibnamefont
  {Witten}},\ }\bibfield  {title} {\bibinfo {title} {{Baryons in the $1/N$
  Expansion}},\ }\href {https://doi.org/10.1016/0550-3213(79)90232-3}
  {\bibfield  {journal} {\bibinfo  {journal} {Nucl. Phys. B}\ }\textbf
  {\bibinfo {volume} {160}},\ \bibinfo {pages} {57} (\bibinfo {year}
  {1979})}\BibitemShut {NoStop}%
\bibitem [{\citenamefont {Prasza\l{}owicz}\ \emph {et~al.}(1995)\citenamefont
  {Prasza\l{}owicz}, \citenamefont {Blotz},\ and\ \citenamefont
  {Goeke}}]{Praszalowicz:1995vi}%
  \BibitemOpen
  \bibfield  {author} {\bibinfo {author} {\bibfnamefont {M.}~\bibnamefont
  {Prasza\l{}owicz}}, \bibinfo {author} {\bibfnamefont {A.}~\bibnamefont
  {Blotz}},\ and\ \bibinfo {author} {\bibfnamefont {K.}~\bibnamefont {Goeke}},\
  }\bibfield  {title} {\bibinfo {title} {{The constituent quark limit and the
  skyrmion limit of chiral quark soliton model}},\ }\href
  {https://doi.org/10.1016/0370-2693(95)00653-3} {\bibfield  {journal}
  {\bibinfo  {journal} {Phys. Lett. B}\ }\textbf {\bibinfo {volume} {354}},\
  \bibinfo {pages} {415} (\bibinfo {year} {1995})},\ \Eprint
  {https://arxiv.org/abs/hep-ph/9505328} {arXiv:hep-ph/9505328} \BibitemShut
  {NoStop}%
\bibitem [{\citenamefont {Christov}\ \emph {et~al.}(1996)\citenamefont
  {Christov}, \citenamefont {Blotz}, \citenamefont {Kim}, \citenamefont
  {Pobylitsa}, \citenamefont {Watabe}, \citenamefont {Meissner}, \citenamefont
  {Ruiz~Arriola},\ and\ \citenamefont {Goeke}}]{Christov:1995vm}%
  \BibitemOpen
  \bibfield  {author} {\bibinfo {author} {\bibfnamefont {C.~V.}\ \bibnamefont
  {Christov}}, \bibinfo {author} {\bibfnamefont {A.}~\bibnamefont {Blotz}},
  \bibinfo {author} {\bibfnamefont {H.-C.}\ \bibnamefont {Kim}}, \bibinfo
  {author} {\bibfnamefont {P.}~\bibnamefont {Pobylitsa}}, \bibinfo {author}
  {\bibfnamefont {T.}~\bibnamefont {Watabe}}, \bibinfo {author} {\bibfnamefont
  {T.}~\bibnamefont {Meissner}}, \bibinfo {author} {\bibfnamefont
  {E.}~\bibnamefont {Ruiz~Arriola}},\ and\ \bibinfo {author} {\bibfnamefont
  {K.}~\bibnamefont {Goeke}},\ }\bibfield  {title} {\bibinfo {title} {{Baryons
  as nontopological chiral solitons}},\ }\href
  {https://doi.org/10.1016/0146-6410(96)00057-9} {\bibfield  {journal}
  {\bibinfo  {journal} {Prog. Part. Nucl. Phys.}\ }\textbf {\bibinfo {volume}
  {37}},\ \bibinfo {pages} {91} (\bibinfo {year} {1996})},\ \Eprint
  {https://arxiv.org/abs/hep-ph/9604441} {arXiv:hep-ph/9604441} \BibitemShut
  {NoStop}%
\bibitem [{\citenamefont {Kim}\ and\ \citenamefont
  {Weiss}(2024)}]{Kim:2023pll}%
  \BibitemOpen
  \bibfield  {author} {\bibinfo {author} {\bibfnamefont {J.-Y.}\ \bibnamefont
  {Kim}}\ and\ \bibinfo {author} {\bibfnamefont {C.}~\bibnamefont {Weiss}},\
  }\bibfield  {title} {\bibinfo {title} {{Instanton effects in twist-3
  generalized parton distributions}},\ }\href
  {https://doi.org/10.1016/j.physletb.2023.138387} {\bibfield  {journal}
  {\bibinfo  {journal} {Phys. Lett. B}\ }\textbf {\bibinfo {volume} {848}},\
  \bibinfo {pages} {138387} (\bibinfo {year} {2024})},\ \Eprint
  {https://arxiv.org/abs/2310.16890} {arXiv:2310.16890 [hep-ph]} \BibitemShut
  {NoStop}%
\bibitem [{\citenamefont {Balla}\ \emph {et~al.}(1998)\citenamefont {Balla},
  \citenamefont {Polyakov},\ and\ \citenamefont {Weiss}}]{Balla:1997hf}%
  \BibitemOpen
  \bibfield  {author} {\bibinfo {author} {\bibfnamefont {J.}~\bibnamefont
  {Balla}}, \bibinfo {author} {\bibfnamefont {M.~V.}\ \bibnamefont
  {Polyakov}},\ and\ \bibinfo {author} {\bibfnamefont {C.}~\bibnamefont
  {Weiss}},\ }\bibfield  {title} {\bibinfo {title} {{Nucleon matrix elements of
  higher twist operators from the instanton vacuum}},\ }\href
  {https://doi.org/10.1016/S0550-3213(98)00638-5} {\bibfield  {journal}
  {\bibinfo  {journal} {Nucl. Phys. B}\ }\textbf {\bibinfo {volume} {510}},\
  \bibinfo {pages} {327} (\bibinfo {year} {1998})},\ \Eprint
  {https://arxiv.org/abs/hep-ph/9707515} {arXiv:hep-ph/9707515} \BibitemShut
  {NoStop}%
\bibitem [{\citenamefont {Lorc\'e}\ \emph {et~al.}(2018)\citenamefont
  {Lorc\'e}, \citenamefont {Mantovani},\ and\ \citenamefont
  {Pasquini}}]{Lorce:2017wkb}%
  \BibitemOpen
  \bibfield  {author} {\bibinfo {author} {\bibfnamefont {C.}~\bibnamefont
  {Lorc\'e}}, \bibinfo {author} {\bibfnamefont {L.}~\bibnamefont {Mantovani}},\
  and\ \bibinfo {author} {\bibfnamefont {B.}~\bibnamefont {Pasquini}},\
  }\bibfield  {title} {\bibinfo {title} {{Spatial distribution of angular
  momentum inside the nucleon}},\ }\href
  {https://doi.org/10.1016/j.physletb.2017.11.018} {\bibfield  {journal}
  {\bibinfo  {journal} {Phys. Lett. B}\ }\textbf {\bibinfo {volume} {776}},\
  \bibinfo {pages} {38} (\bibinfo {year} {2018})},\ \Eprint
  {https://arxiv.org/abs/1704.08557} {arXiv:1704.08557 [hep-ph]} \BibitemShut
  {NoStop}%
\bibitem [{\citenamefont {Kim}\ \emph {et~al.}(2023{\natexlab{a}})\citenamefont
  {Kim}, \citenamefont {Won}, \citenamefont {Goity},\ and\ \citenamefont
  {Weiss}}]{Kim:2023xvw}%
  \BibitemOpen
  \bibfield  {author} {\bibinfo {author} {\bibfnamefont {J.-Y.}\ \bibnamefont
  {Kim}}, \bibinfo {author} {\bibfnamefont {H.-Y.}\ \bibnamefont {Won}},
  \bibinfo {author} {\bibfnamefont {J.~L.}\ \bibnamefont {Goity}},\ and\
  \bibinfo {author} {\bibfnamefont {C.}~\bibnamefont {Weiss}},\ }\bibfield
  {title} {\bibinfo {title} {{QCD angular momentum in $N \rightarrow \Delta$
  transitions}},\ }\href {https://doi.org/10.1016/j.physletb.2023.138083}
  {\bibfield  {journal} {\bibinfo  {journal} {Phys. Lett. B}\ }\textbf
  {\bibinfo {volume} {844}},\ \bibinfo {pages} {138083} (\bibinfo {year}
  {2023}{\natexlab{a}})},\ \Eprint {https://arxiv.org/abs/2304.08575}
  {arXiv:2304.08575 [hep-ph]} \BibitemShut {NoStop}%
\bibitem [{\citenamefont {Berestetskii}\ \emph {et~al.}(1982)\citenamefont
  {Berestetskii}, \citenamefont {Lifshitz},\ and\ \citenamefont
  {Pitaevskii}}]{Berestetskii:1982qgu}%
  \BibitemOpen
  \bibfield  {author} {\bibinfo {author} {\bibfnamefont {V.~B.}\ \bibnamefont
  {Berestetskii}}, \bibinfo {author} {\bibfnamefont {E.~M.}\ \bibnamefont
  {Lifshitz}},\ and\ \bibinfo {author} {\bibfnamefont {L.~P.}\ \bibnamefont
  {Pitaevskii}},\ }\href@noop {} {\emph {\bibinfo {title} {{Quantum
  Electrodynamics}}}},\ \bibinfo {series} {Course of Theoretical Physics},
  Vol.~\bibinfo {volume} {4}\ (\bibinfo  {publisher} {Pergamon Press},\
  \bibinfo {address} {Oxford},\ \bibinfo {year} {1982})\BibitemShut {NoStop}%
\bibitem [{\citenamefont {Shuryak}(1989)}]{Shuryak:1988rf}%
  \BibitemOpen
  \bibfield  {author} {\bibinfo {author} {\bibfnamefont {E.~V.}\ \bibnamefont
  {Shuryak}},\ }\bibfield  {title} {\bibinfo {title} {{Instantons in {QCD}. 1.
  Properties of the ``Instanton Liquid''}},\ }\href
  {https://doi.org/10.1016/0550-3213(89)90618-4} {\bibfield  {journal}
  {\bibinfo  {journal} {Nucl. Phys. B}\ }\textbf {\bibinfo {volume} {319}},\
  \bibinfo {pages} {521} (\bibinfo {year} {1989})}\BibitemShut {NoStop}%
\bibitem [{\citenamefont {Diakonov}\ and\ \citenamefont
  {Eides}(1983)}]{Diakonov:1983bny}%
  \BibitemOpen
  \bibfield  {author} {\bibinfo {author} {\bibfnamefont {D.}~\bibnamefont
  {Diakonov}}\ and\ \bibinfo {author} {\bibfnamefont {M.~I.}\ \bibnamefont
  {Eides}},\ }\bibfield  {title} {\bibinfo {title} {{Chiral Lagrangian from a
  functional integral over quarks}},\ }\href@noop {} {\bibfield  {journal}
  {\bibinfo  {journal} {JETP Lett.}\ }\textbf {\bibinfo {volume} {38}},\
  \bibinfo {pages} {433} (\bibinfo {year} {1983})}\BibitemShut {NoStop}%
\bibitem [{\citenamefont {Weinberg}(1968)}]{Weinberg:1968de}%
  \BibitemOpen
  \bibfield  {author} {\bibinfo {author} {\bibfnamefont {S.}~\bibnamefont
  {Weinberg}},\ }\bibfield  {title} {\bibinfo {title} {{Nonlinear realizations
  of chiral symmetry}},\ }\href {https://doi.org/10.1103/PhysRev.166.1568}
  {\bibfield  {journal} {\bibinfo  {journal} {Phys. Rev.}\ }\textbf {\bibinfo
  {volume} {166}},\ \bibinfo {pages} {1568} (\bibinfo {year}
  {1968})}\BibitemShut {NoStop}%
\bibitem [{\citenamefont {Gasser}\ and\ \citenamefont
  {Leutwyler}(1984)}]{Gasser:1983yg}%
  \BibitemOpen
  \bibfield  {author} {\bibinfo {author} {\bibfnamefont {J.}~\bibnamefont
  {Gasser}}\ and\ \bibinfo {author} {\bibfnamefont {H.}~\bibnamefont
  {Leutwyler}},\ }\bibfield  {title} {\bibinfo {title} {{Chiral Perturbation
  Theory to One Loop}},\ }\href {https://doi.org/10.1016/0003-4916(84)90242-2}
  {\bibfield  {journal} {\bibinfo  {journal} {Annals Phys.}\ }\textbf {\bibinfo
  {volume} {158}},\ \bibinfo {pages} {142} (\bibinfo {year}
  {1984})}\BibitemShut {NoStop}%
\bibitem [{\citenamefont {Skyrme}(1961)}]{Skyrme:1961vq}%
  \BibitemOpen
  \bibfield  {author} {\bibinfo {author} {\bibfnamefont {T.~H.~R.}\
  \bibnamefont {Skyrme}},\ }\bibfield  {title} {\bibinfo {title} {{A Nonlinear
  field theory}},\ }\href {https://doi.org/10.1098/rspa.1961.0018} {\bibfield
  {journal} {\bibinfo  {journal} {Proc. Roy. Soc. Lond. A}\ }\textbf {\bibinfo
  {volume} {260}},\ \bibinfo {pages} {127} (\bibinfo {year}
  {1961})}\BibitemShut {NoStop}%
\bibitem [{\citenamefont {Adkins}\ \emph {et~al.}(1983)\citenamefont {Adkins},
  \citenamefont {Nappi},\ and\ \citenamefont {Witten}}]{Adkins:1983ya}%
  \BibitemOpen
  \bibfield  {author} {\bibinfo {author} {\bibfnamefont {G.~S.}\ \bibnamefont
  {Adkins}}, \bibinfo {author} {\bibfnamefont {C.~R.}\ \bibnamefont {Nappi}},\
  and\ \bibinfo {author} {\bibfnamefont {E.}~\bibnamefont {Witten}},\
  }\bibfield  {title} {\bibinfo {title} {{Static Properties of Nucleons in the
  Skyrme Model}},\ }\href {https://doi.org/10.1016/0550-3213(83)90559-X}
  {\bibfield  {journal} {\bibinfo  {journal} {Nucl. Phys. B}\ }\textbf
  {\bibinfo {volume} {228}},\ \bibinfo {pages} {552} (\bibinfo {year}
  {1983})}\BibitemShut {NoStop}%
\bibitem [{\citenamefont {Zahed}\ and\ \citenamefont
  {Brown}(1986)}]{Zahed:1986qz}%
  \BibitemOpen
  \bibfield  {author} {\bibinfo {author} {\bibfnamefont {I.}~\bibnamefont
  {Zahed}}\ and\ \bibinfo {author} {\bibfnamefont {G.~E.}\ \bibnamefont
  {Brown}},\ }\bibfield  {title} {\bibinfo {title} {{The Skyrme Model}},\
  }\href {https://doi.org/10.1016/0370-1573(86)90142-0} {\bibfield  {journal}
  {\bibinfo  {journal} {Phys. Rept.}\ }\textbf {\bibinfo {volume} {142}},\
  \bibinfo {pages} {1} (\bibinfo {year} {1986})}\BibitemShut {NoStop}%
\bibitem [{\citenamefont {Choi}\ and\ \citenamefont {Kim}(2004)}]{Choi:2003cz}%
  \BibitemOpen
  \bibfield  {author} {\bibinfo {author} {\bibfnamefont {H.-A.}\ \bibnamefont
  {Choi}}\ and\ \bibinfo {author} {\bibfnamefont {H.-C.}\ \bibnamefont {Kim}},\
  }\bibfield  {title} {\bibinfo {title} {{Effective chiral Lagrangian in the
  chiral limit from the instanton vacuum}},\ }\href
  {https://doi.org/10.1103/PhysRevD.69.054004} {\bibfield  {journal} {\bibinfo
  {journal} {Phys. Rev. D}\ }\textbf {\bibinfo {volume} {69}},\ \bibinfo
  {pages} {054004} (\bibinfo {year} {2004})},\ \Eprint
  {https://arxiv.org/abs/hep-ph/0308171} {arXiv:hep-ph/0308171} \BibitemShut
  {NoStop}%
\bibitem [{\citenamefont {Goeke}\ \emph {et~al.}(2007)\citenamefont {Goeke},
  \citenamefont {Musakhanov},\ and\ \citenamefont {Siddikov}}]{Goeke:2007bj}%
  \BibitemOpen
  \bibfield  {author} {\bibinfo {author} {\bibfnamefont {K.}~\bibnamefont
  {Goeke}}, \bibinfo {author} {\bibfnamefont {M.~M.}\ \bibnamefont
  {Musakhanov}},\ and\ \bibinfo {author} {\bibfnamefont {M.}~\bibnamefont
  {Siddikov}},\ }\bibfield  {title} {\bibinfo {title} {{Low energy constants of
  $\chi$PT from the instanton vacuum model}},\ }\href
  {https://doi.org/10.1103/PhysRevD.76.076007} {\bibfield  {journal} {\bibinfo
  {journal} {Phys. Rev. D}\ }\textbf {\bibinfo {volume} {76}},\ \bibinfo
  {pages} {076007} (\bibinfo {year} {2007})},\ \Eprint
  {https://arxiv.org/abs/0707.1997} {arXiv:0707.1997 [hep-ph]} \BibitemShut
  {NoStop}%
\bibitem [{\citenamefont {Wakamatsu}\ and\ \citenamefont
  {Yoshiki}(1991)}]{Wakamatsu:1990ud}%
  \BibitemOpen
  \bibfield  {author} {\bibinfo {author} {\bibfnamefont {M.}~\bibnamefont
  {Wakamatsu}}\ and\ \bibinfo {author} {\bibfnamefont {H.}~\bibnamefont
  {Yoshiki}},\ }\bibfield  {title} {\bibinfo {title} {{A chiral quark model of
  the nucleon}},\ }\href {https://doi.org/10.1016/0375-9474(91)90263-6}
  {\bibfield  {journal} {\bibinfo  {journal} {Nucl. Phys. A}\ }\textbf
  {\bibinfo {volume} {524}},\ \bibinfo {pages} {561} (\bibinfo {year}
  {1991})}\BibitemShut {NoStop}%
\bibitem [{\citenamefont {Diakonov}\ \emph
  {et~al.}(1996{\natexlab{b}})\citenamefont {Diakonov}, \citenamefont {Petrov},
  \citenamefont {Pobylitsa}, \citenamefont {Polyakov},\ and\ \citenamefont
  {Weiss}}]{Diakonov:1996sr}%
  \BibitemOpen
  \bibfield  {author} {\bibinfo {author} {\bibfnamefont {D.}~\bibnamefont
  {Diakonov}}, \bibinfo {author} {\bibfnamefont {V.}~\bibnamefont {Petrov}},
  \bibinfo {author} {\bibfnamefont {P.}~\bibnamefont {Pobylitsa}}, \bibinfo
  {author} {\bibfnamefont {M.~V.}\ \bibnamefont {Polyakov}},\ and\ \bibinfo
  {author} {\bibfnamefont {C.}~\bibnamefont {Weiss}},\ }\bibfield  {title}
  {\bibinfo {title} {{Nucleon parton distributions at low normalization point
  in the large $N_c$ limit}},\ }\href
  {https://doi.org/10.1016/S0550-3213(96)00486-5} {\bibfield  {journal}
  {\bibinfo  {journal} {Nucl. Phys. B}\ }\textbf {\bibinfo {volume} {480}},\
  \bibinfo {pages} {341} (\bibinfo {year} {1996}{\natexlab{b}})},\ \Eprint
  {https://arxiv.org/abs/hep-ph/9606314} {arXiv:hep-ph/9606314} \BibitemShut
  {NoStop}%
\bibitem [{\citenamefont {Diakonov}\ \emph {et~al.}(1997)\citenamefont
  {Diakonov}, \citenamefont {Petrov}, \citenamefont {Pobylitsa}, \citenamefont
  {Polyakov},\ and\ \citenamefont {Weiss}}]{Diakonov:1997vc}%
  \BibitemOpen
  \bibfield  {author} {\bibinfo {author} {\bibfnamefont {D.}~\bibnamefont
  {Diakonov}}, \bibinfo {author} {\bibfnamefont {V.~Y.}\ \bibnamefont
  {Petrov}}, \bibinfo {author} {\bibfnamefont {P.~V.}\ \bibnamefont
  {Pobylitsa}}, \bibinfo {author} {\bibfnamefont {M.~V.}\ \bibnamefont
  {Polyakov}},\ and\ \bibinfo {author} {\bibfnamefont {C.}~\bibnamefont
  {Weiss}},\ }\bibfield  {title} {\bibinfo {title} {{Unpolarized and polarized
  quark distributions in the large $N_c$ limit}},\ }\href
  {https://doi.org/10.1103/PhysRevD.56.4069} {\bibfield  {journal} {\bibinfo
  {journal} {Phys. Rev. D}\ }\textbf {\bibinfo {volume} {56}},\ \bibinfo
  {pages} {4069} (\bibinfo {year} {1997})},\ \Eprint
  {https://arxiv.org/abs/hep-ph/9703420} {arXiv:hep-ph/9703420} \BibitemShut
  {NoStop}%
\bibitem [{\citenamefont {Kahana}\ \emph {et~al.}(1984)\citenamefont {Kahana},
  \citenamefont {Ripka},\ and\ \citenamefont {Soni}}]{Kahana:1984dx}%
  \BibitemOpen
  \bibfield  {author} {\bibinfo {author} {\bibfnamefont {S.}~\bibnamefont
  {Kahana}}, \bibinfo {author} {\bibfnamefont {G.}~\bibnamefont {Ripka}},\ and\
  \bibinfo {author} {\bibfnamefont {V.}~\bibnamefont {Soni}},\ }\bibfield
  {title} {\bibinfo {title} {{Soliton with Valence Quarks in the Chiral
  Invariant Sigma Model}},\ }\href
  {https://doi.org/10.1016/0375-9474(84)90306-3} {\bibfield  {journal}
  {\bibinfo  {journal} {Nucl. Phys. A}\ }\textbf {\bibinfo {volume} {415}},\
  \bibinfo {pages} {351} (\bibinfo {year} {1984})}\BibitemShut {NoStop}%
\bibitem [{\citenamefont {Kahana}\ and\ \citenamefont
  {Ripka}(1984)}]{Kahana:1984be}%
  \BibitemOpen
  \bibfield  {author} {\bibinfo {author} {\bibfnamefont {S.}~\bibnamefont
  {Kahana}}\ and\ \bibinfo {author} {\bibfnamefont {G.}~\bibnamefont {Ripka}},\
  }\bibfield  {title} {\bibinfo {title} {{Baryon Density of Quarks Coupled to a
  Chiral Field}},\ }\href {https://doi.org/10.1016/0375-9474(84)90692-4}
  {\bibfield  {journal} {\bibinfo  {journal} {Nucl. Phys. A}\ }\textbf
  {\bibinfo {volume} {429}},\ \bibinfo {pages} {462} (\bibinfo {year}
  {1984})}\BibitemShut {NoStop}%
\bibitem [{\citenamefont {Schweitzer}\ \emph {et~al.}(2013)\citenamefont
  {Schweitzer}, \citenamefont {Strikman},\ and\ \citenamefont
  {Weiss}}]{Schweitzer:2012hh}%
  \BibitemOpen
  \bibfield  {author} {\bibinfo {author} {\bibfnamefont {P.}~\bibnamefont
  {Schweitzer}}, \bibinfo {author} {\bibfnamefont {M.}~\bibnamefont
  {Strikman}},\ and\ \bibinfo {author} {\bibfnamefont {C.}~\bibnamefont
  {Weiss}},\ }\bibfield  {title} {\bibinfo {title} {{Intrinsic transverse
  momentum and parton correlations from dynamical chiral symmetry breaking}},\
  }\href {https://doi.org/10.1007/JHEP01(2013)163} {\bibfield  {journal}
  {\bibinfo  {journal} {JHEP}\ }\textbf {\bibinfo {volume} {01}},\ \bibinfo
  {pages} {163}},\ \Eprint {https://arxiv.org/abs/1210.1267} {arXiv:1210.1267
  [hep-ph]} \BibitemShut {NoStop}%
\bibitem [{\citenamefont {Landau}\ and\ \citenamefont
  {Lifshits}(1991)}]{Landau:1991wop}%
  \BibitemOpen
  \bibfield  {author} {\bibinfo {author} {\bibfnamefont {L.~D.}\ \bibnamefont
  {Landau}}\ and\ \bibinfo {author} {\bibfnamefont {E.~M.}\ \bibnamefont
  {Lifshits}},\ }\href@noop {} {\emph {\bibinfo {title} {{Quantum Mechanics}:
  {Non-Relativistic Theory}}}},\ \bibinfo {series} {Course of Theoretical
  Physics}, Vol.\ \bibinfo {volume} {v.3}\ (\bibinfo  {publisher}
  {Butterworth-Heinemann},\ \bibinfo {address} {Oxford},\ \bibinfo {year}
  {1991})\BibitemShut {NoStop}%
\bibitem [{\citenamefont {Kim}\ \emph {et~al.}(2023{\natexlab{b}})\citenamefont
  {Kim}, \citenamefont {Won}, \citenamefont {Kim}, \citenamefont {Goity},\ and\
  \citenamefont {Weiss}}]{kk}%
  \BibitemOpen
  \bibfield  {author} {\bibinfo {author} {\bibfnamefont {J.-Y.}\ \bibnamefont
  {Kim}}, \bibinfo {author} {\bibfnamefont {H.-Y.}\ \bibnamefont {Won}},
  \bibinfo {author} {\bibfnamefont {H.-C.}\ \bibnamefont {Kim}}, \bibinfo
  {author} {\bibfnamefont {J.~L.}\ \bibnamefont {Goity}},\ and\ \bibinfo
  {author} {\bibfnamefont {C.}~\bibnamefont {Weiss}},\ }\href@noop {}
  {}\bibinfo {howpublished} {to appear} (\bibinfo {year}
  {2023}{\natexlab{b}})\BibitemShut {NoStop}%
\bibitem [{\citenamefont {Lorc\'e}\ and\ \citenamefont
  {Pasquini}(2011)}]{Lorce:2011kd}%
  \BibitemOpen
  \bibfield  {author} {\bibinfo {author} {\bibfnamefont {C.}~\bibnamefont
  {Lorc\'e}}\ and\ \bibinfo {author} {\bibfnamefont {B.}~\bibnamefont
  {Pasquini}},\ }\bibfield  {title} {\bibinfo {title} {{Quark Wigner
  Distributions and Orbital Angular Momentum}},\ }\href
  {https://doi.org/10.1103/PhysRevD.84.014015} {\bibfield  {journal} {\bibinfo
  {journal} {Phys. Rev. D}\ }\textbf {\bibinfo {volume} {84}},\ \bibinfo
  {pages} {014015} (\bibinfo {year} {2011})},\ \Eprint
  {https://arxiv.org/abs/1106.0139} {arXiv:1106.0139 [hep-ph]} \BibitemShut
  {NoStop}%
\bibitem [{\citenamefont {Kim}\ \emph {et~al.}(1996)\citenamefont {Kim},
  \citenamefont {Polyakov},\ and\ \citenamefont {Goeke}}]{Kim:1995bq}%
  \BibitemOpen
  \bibfield  {author} {\bibinfo {author} {\bibfnamefont {H.-C.}\ \bibnamefont
  {Kim}}, \bibinfo {author} {\bibfnamefont {M.~V.}\ \bibnamefont {Polyakov}},\
  and\ \bibinfo {author} {\bibfnamefont {K.}~\bibnamefont {Goeke}},\ }\bibfield
   {title} {\bibinfo {title} {{Nucleon tensor charges in the SU(2) chiral
  quark-soliton model}},\ }\href {https://doi.org/10.1103/PhysRevD.53.R4715}
  {\bibfield  {journal} {\bibinfo  {journal} {Phys. Rev. D}\ }\textbf {\bibinfo
  {volume} {53}},\ \bibinfo {pages} {4715} (\bibinfo {year} {1996})},\ \Eprint
  {https://arxiv.org/abs/hep-ph/9509283} {arXiv:hep-ph/9509283} \BibitemShut
  {NoStop}%
\end{thebibliography}%
\end{document}